# Artificial Intelligence based Load balancing in SDN: A Comprehensive Survey

Ahmed Hazim Alhilali Ahmadreza Montazerolghaem

*Faculty of Computer Engineering, University of Isfahan, Isfahan, Iran*

**ABSTRACT**

In the future, it is anticipated that software-defined networking (SDN) will become the preferred platform for deploying diverse networks. Compared to traditional networks, SDN separates the control and data planes for efficient domain-wide traffic routing and management. The controllers in the control plane are responsible for programming data plane forwarding devices, while the top layer, the application plane, enforces policies and programs the network. The different levels of the SDN use interfaces for communication. However, SDN faces challenges with traffic distribution, such as load imbalance, which can negatively affect the network performance. Consequently, developers have developed various SDN load-balancing solutions to enhance SDN effectiveness. In addition, researchers are considering the potential of implementing some artificial intelligence (AI) approaches into SDN to improve network resource usage and overall performance due to the fast growth of the AI field. This survey focuses on the following: Firstly, analyzing the SDN architecture and investigating the problem of load balancing in SDN. Secondly, categorizing AI-based load balancing methods and thoroughly assessing these mechanisms from various perspectives, such as the algorithm/technique employed, the tackled problem, and their strengths and weaknesses. Thirdly, summarizing the metrics utilized to measure the effectiveness of these techniques. Finally, identifying the trends and challenges of AI-based load balancing for future research.

## 1. INTRODUCTION

In recent years, network requirements are changing quickly as network traffic and quality conditions are growing, putting more pressure on the network infrastructure. Traditional network topologies still struggle to adapt to the dynamic nature of modern networks because of their inflexibility. Developers have developed the concept of "Software-Defined Networking" (SDN) to address the need for flexible networks. The concept of SDN was initially proposed by researchers at Stanford University [1]. Service providers have confidence in SDN because it can efficiently manage most network components' functions. SDN provides a network design that separates the control plane from the data plane and allows for a more flexible, scalable, and cost-effective network architecture [2]. A centralized SDN controller is part of the control plane and is responsible for routing packets [3]. At the same time, the data plane is the infrastructural layer, which comprises interconnected forwarding units, such as software-defined networking (SDN) switches. In order to properly apply SDN-based technologies, the networking components need to include the software in their physical infrastructure [4].

Critical technologies supported by the SDN are OpenFlow and Path Computation Element [5]. The Open Networking Foundation (ONF) strongly advises using OpenFlow because it is the standard protocol that decouples the control plane from the switch and offers a communication link between the SDN layers (control and data layers) [5]. Internet Engineering Task Force (IETF) [6] supports Path Calculation Element (PCE) for closed settings like data centers where path computation is transferred to the controller. The OpenFlow protocol development makes network traffic monitoring effective and efficient and provides flexible topology [5]. It allows the software to operate on various routers and promotes packet path association across the network. As conventional networks are incapable of providing a global view of network structure and resources, they did not discuss load-balancing techniques in detail previously. Since the controller provides information about the network resources that could be used for optimizing the load, it is an ideal environment for load balancing implementation.

Load balancing (LB) is a strategy whereby numerous resources are used to handle a single task to prevent network overload [7]. LB generally aims to minimize the throughput and response time and optimize the network traffic. In conventional networks, load balancing strategies are notoriously inaccurate, while in SDN, it is characterized by its accuracy and high performance.

The comprehensive study of SDN can be challenging because of its multidimensional nature. Although load balancing can improve SDN performance, more studies need to be on it, prompting the authors to conduct further investigation in the area of load balancing in SDN. To our knowledge, this is the first comprehensive Load Balancing (LB) survey in software-defined networking that concentrates on the existing Artificial Intelligence (AI) techniques and their effects on SDN performance. Even though there have been several in-depth studies on SDN LB, such as Ahmad and Khan [8] (2018) Gebremariam et al. [9] (2019), Belgaum et al. [10] (2021), Latah and Toker [11] (2019), Hota et al. [12] (2019), Belgaum et al. [13] (2020) our work relies on different aspects to provide a new classification and analysis for LB methods. In this paper, we examine the load-balancing strategies, policies, and algorithms currently used in SDN, study the variables that affect the load distribution and evaluate its effectiveness, and discuss the significant trends and challenges in SDN load balancing that can help researchers to improve SDN performance. A study by Haris and Khan [8] (2018) offered a systematic study of current techniques and tools for load balancing in cloud computing. In this regard, some important criteria like as throughput, scalability, fault tolerance, and reaction time are taken into account in the evaluation. However, this paper has ignored the published articles between 2016 and 2018. Also, Gebremariam et al. [9] (2019) provided a comprehensive overview of the core AI/ML application fields in SDN and Network functions virtualization (NFV)-based networks. The survey classified essential advancements in these fields according to their application trend and determined the AI methodologies used. However, none of this research considered the load balancing aspects in software-defined networks. Furthermore, the objective of Belgaum et al. [10] (2021) is to study two artificial intelligence optimization approaches, including Ant Colony Optimization (ACO) and Particle Swarm Optimization (PSO), and their application for load balancing in Software Defined Networking (SDN). It suggested incorporating a reliable link and node selection approach to enhance the network performance and improve the load. In contrast, in Latah and Toker [11] (2019), three distinct sub-disciplines of Artificial Intelligence (AI) have been investigated: machine learning, meta-heuristics, and fuzzy inference systems. The work highlights the application areas of AI-based techniques and their improvements in the SDN paradigm. However, the drawback associated with the mentioned studies is that it is considered only some of the available AI methods. Hota et al. [12] (2019) suggested a literature review of load balancing algorithms in cloud computing. The algorithms have been categorized into three groups, namely, metaheuristic, heuristic, and hybrid, based on their adopted algorithms. The advantages, disadvantages, and optimization techniques of each algorithm have been outlined. Nevertheless, the study has not taken into account any recently published papers. Finally, Belgaum et al. [13] (2020) suggested a methodical investigation of load balancing techniques and algorithms used by different researchers. Depending on the strategy used to address SDN load balancing difficulties, the articles were divided into two groups: artificial intelligence-based techniques and classical load balancing-based approaches. Similarly, this work focused on the problems that have been raised, the strategies employed, and solutions suggested. The authors observed that several techniques did not fulfill specific crucial requirements necessary to enhance the efficiency of the existing SDN load balancing methods.

Table 1 presents a comparison between our study and previous surveys based on several aspects such as review type, publication year, classification, main topic, future work, and years of reviewed papers. The comparison highlights that only two papers provide comprehensive reviews of both dynamic and static load balancing methods. As a result, our research is the first to concentrate on the impact of existing Artificial Intelligence (AI) based load balancing techniques on SDN performance in the domain of software-defined networking.

The main contributions of this work as follow:

- We comprehensively survey the various Artificial Intelligence (AI) approaches to address load balancing problems and their impacts on software-defined network (SDN) performance.
- We present a detailed evaluation and categorization of the existing AI-based LB mechanisms while highlighting their primary features, including the algorithm or technique, the addressed problem, and the strength and weaknesses of each methodology.
- We introduce the most used parameters to assess the effectiveness of the proposed techniques.
- Finally, this survey highlights the trends and challenges that need to be addressed in the future as prospects for further research. These prospects could provide researchers with assistance and inspiration for future SDN LB endeavors.

The rest of this article is structured as follows. In section 2, the background of SDN architecture, the key benefits of utilizing SDN, and the concept and structure of load balancing are presented. Section 3 reviews the chosen load balancing methods and classifies them into four categories. Section 4 provides a comparison of the results obtained from these techniques. Section 5 outlines the trends and challenges. Finally, in section 6, the research is concluded.

*Table 1: Related studies in the field of load balancing*

| Authors | Review type | Publication year | Classification | Main topic | Future work | Years of review papers |
|---|---|---|---|---|---|---|
| Ahmad and Khan [8] | Systematic Literature Review (SLR) | 2018 | No | Cloud | Not presented | 2010-2015 |
| Gebremariam et al. [9] | Survey | 2019 | Yes | SDN and NFV | Presented | 2016-2018 |
| Belgaum et al. [10] | Survey | 2021 | No | SDN | Presented | 2016-2019 |
| Latah and Toker [11] | Comprehensive overview | 2019 | Yes | SDN | Not presented | 1985-2019 |
| Hota et al. [12] | Comprehensive Review | 2019 | No | Cloud | Not presented | 2008-2016 |
| Belgaum et al. [13] | Systematic Review | 2020 | Yes | SDN | Presented | 2015-2019 |
| Our work | Comprehensive Survey | -------- | Yes | SDN | Presented | 2017-2023 |

## 2. BACKGROUND

This section presents a brief overview about the architecture of SDN and the primary advantages of utilizing SDN. Furthermore, the concept and structure of load balancing have been discussed.

### 2.1. SDN ARCHITECTURE

SDN architecture represents one of the innovative network designs. It provides a central controller that manages the entire network infrastructure. The OpenFlow protocol is best suited for implementing SDN architecture. Compared to traditional networks, SDN architecture combined with the OpenFlow protocol provides network operators with a superior technique for processing flows via controllers. In a conventional network, the control and data plane are integrated into the equipment. In contrast, SDN is an architecture that divides the networks into a control plane and a data plane (forwarding plane). The control plane, which typically comprises one or more controllers, is the network's brain and controls the whole structure. While the real network hardware, such as routers, switches, and middle boxes, that is in charge of transmitting data is represented by the data plane [14].

SDN architecture is organized into three principal planes based on the Open Networking Foundation (ONF). The architecture is depicted in Figure 1.

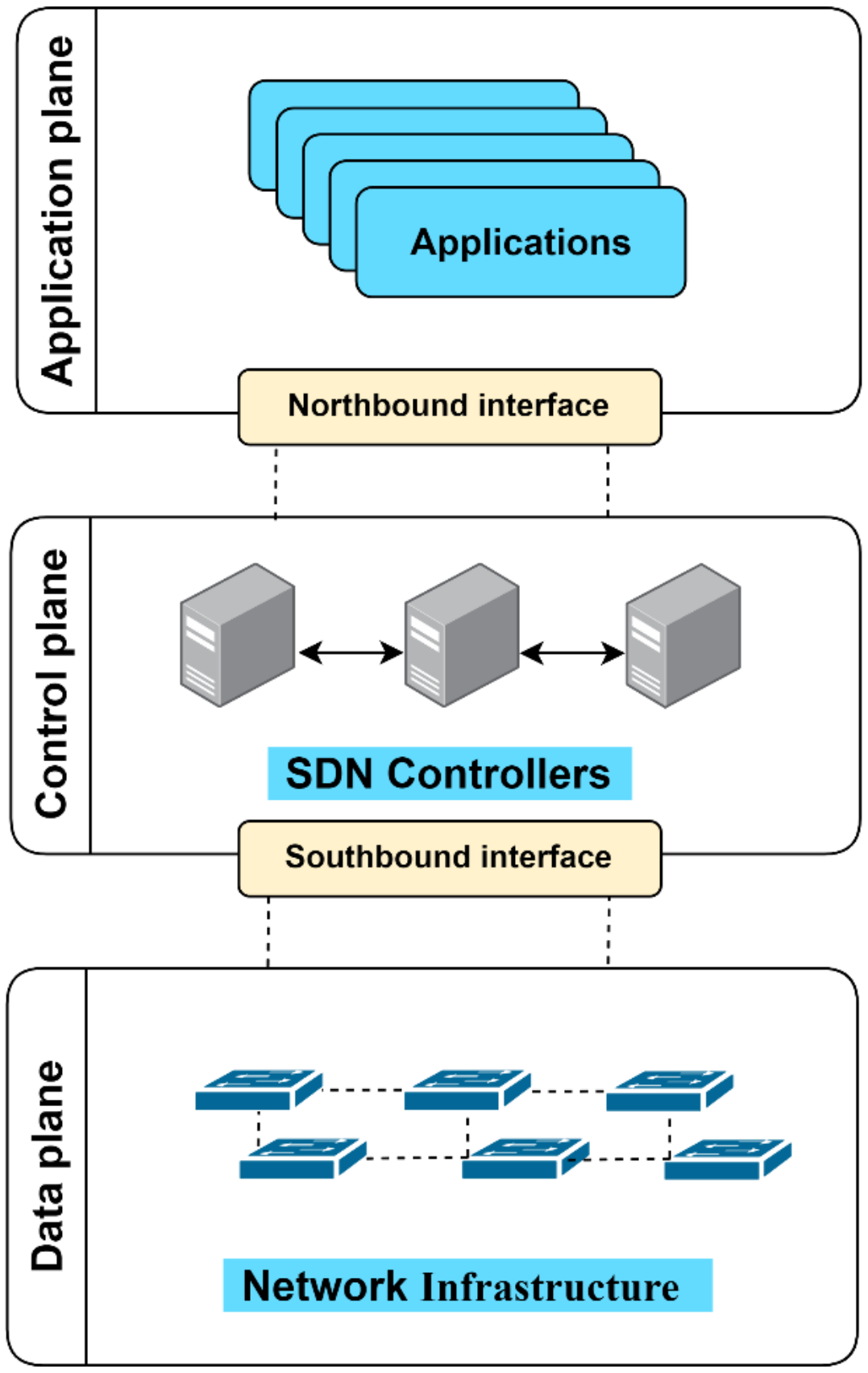

*Figure 1: SDN architecture*

- **Data plane:** It represents the bottom layer in the SDN topology and is considered the network infrastructure. The network forwarding equipment included in this layer, such as routers, physical/virtual switches, access points etc. The main job that is carried out in this layer is packet forwarding in accordance with predetermined guidelines. These rules are defined and installed on the flow table of switches by the SDN controller [14][15].
- **Control plane:** Represents the intermediate layer in an SDN architecture that provides control functionality through software-based SDN controller(s) to manage the network forwarding behavior. The controller manages the network switches, which are responsible for transmitting packets according to specific instructions. Also, it creates an abstract and centralized underlying infrastructure vision for the higher layer [14][16]. The controller uses the southbound API combined with OpenFlow protocol to communicate with the network devices. A part of the SDN controller called the load balancer, which is located in the logical central decision point, used to apply the load lancing algorithms [16][17].
- **Application plane:** This layer includes one or more end-user applications which use the abstract and centralized underlying infrastructure vision to demonstrate their internal decision-making process. In the application implementation process, the programmers use the northbound API to communicate with the SDN controller. This API serves as a software bridge between SDN applications operating on the network and controller platform components [14][15].

### 2.2. SDN ADVANTAGES

SDN provides several advantages to overcoming the challenges presented by conventional network architectures. One of the essential benefits is network programmability. This feature allowed the organizations to have programmatic control over their networks and to grow such networks without affecting performance, reliability, or the quality of the user experience. The SDN eliminates the infrastructure layer complexity and adds visibility for services and applications, thus simplifying network management operations. Network administrators are not required to use custom policies and protocols for the network devices individually in SDN architecture. Simultaneously, an independent controller that is not part of the actual network hardware carries out the control-plane operations. Using SDN enables the network operators to avoid congestion and reduces the complexity of traffic engineering [2].

The scalability issues are significant for Data centers, especially as the number of virtual machines (VMs) grows and they move from one place to another. Therefore, SDN network virtualization presents a significant chance for large-scale data centers. This functionality allowed the network administrators to run Layer 2 traffic across Layer 3 overlays and isolate the MACs of the infrastructure layer devices, making it easier to transfer and create virtual network machines. Moreover, service providers can use the SDN to combine all the network components, such as servers, facilities and clouds, whether physical or virtual, into a single logical network. Consequently, every consumer will have their own personal perspective of the service provider [2][14]. Network device configuration and troubleshooting with SDN can be accomplished through a single controller; thus, it became easy to add and configure devices when needed to grow the network. By offering a programmable platform, SND encourages those interested in networks to use new protocols and ideas and test them in this environment [1][2][14].

### 2.3. LOAD BALANCING MECHANISMS IN SDN

LB technologies are typically employed to enhance the overall performance of distributed systems by effectively spreading incoming clients, requests, and jobs among the available network resources [15]. This technique can be implemented programmatically or in physical equipment to improve the response time, boost throughput, and keep the network from being overloaded. . Integrating the SDN architecture with virtual resources has the potential to enhance energy efficiency and optimize load distribution in the Internet of Multimedia Things (IoMT) [18]. There are many ways to apply load balancing mechanisms, such as static, dynamic or a combination of both [19]. The static methods depend on the system's preliminary information essentially. Static LB mechanisms might be ineffective for all networks due to unexpected user behavior and the immutable load balancer rules. On the other hand, dynamic methods can distribute loads more efficiently than static methods because they use load balancers' pre-programmed patterns [20].

A proper load-balancing approach could effectively reduce response time and packet loss ratio, improve resource utilization, and overload. In addition to this, it has the potential to boost scalability, reliability, the packet delivery ratio, and the longevity of the network. The load-balancing methods need to be analyzed and compared to determine the most effective solution to the load-balancing problem and identify each mechanism's benefits and drawbacks [20]. Different parameters, known as qualitative parameters, such as latency, energy consumption, packet delivery ratio, scalability, etc., should be considered during the comparison process to ensure reliable results [19].

## 3. REVIEW OF SDN LOAD BALANCING BASED ON ARTIFICIAL INTELLIGENCE

These techniques apply an approach known as a meta-heuristic to address real-world challenges. Artificial intelligence (AI) encompasses a variety of topics, including neural networks, natural language processing, deep learning and the AI-based decision-making approaches such as search, planning, and decision theory. In software-defined networking (SDN), load balancing approaches based on artificial intelligence improve learning capabilities and stimulate decision-making.

In this section, different mechanisms that researchers propose will be revised in terms of implementation and evaluation metrics. Also, the paper lists the employed load-balancing strategies' characteristics, including the algorithm or technique, the addressed problem, and the strength and weaknesses of each methodology. Moreover, existing SDN load balancing solutions are classified into four main categories, each with sub-categories based on the used technology. Finally, the section will include an explanation of each technique's application. Figure 2 shows the SDN LB four main categories and its sub-categories.

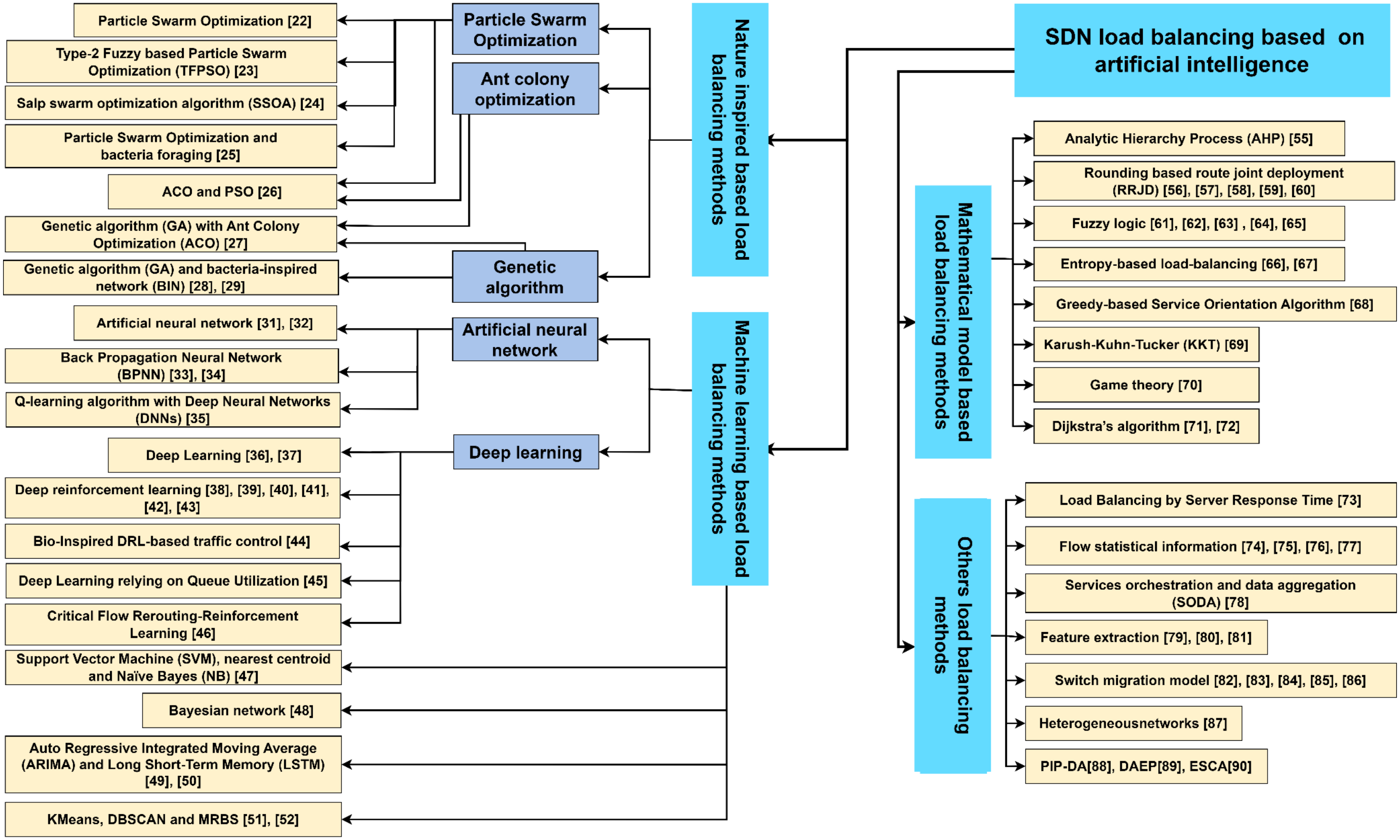

*Figure 2: Classification of AI-based SDN load balancing methods*

### 3.1. SDN LB ALGORITHMS MECHANISM

This section reviews the LB algorithms implementation and classifies the proposed techniques into four main categories based on the nature of the used algorithm. Figure 3 (a) shows the reviewed articles distribution based on the year of publication from 2017 until 2022, and figure 3 (b) shows the number of works for each category.

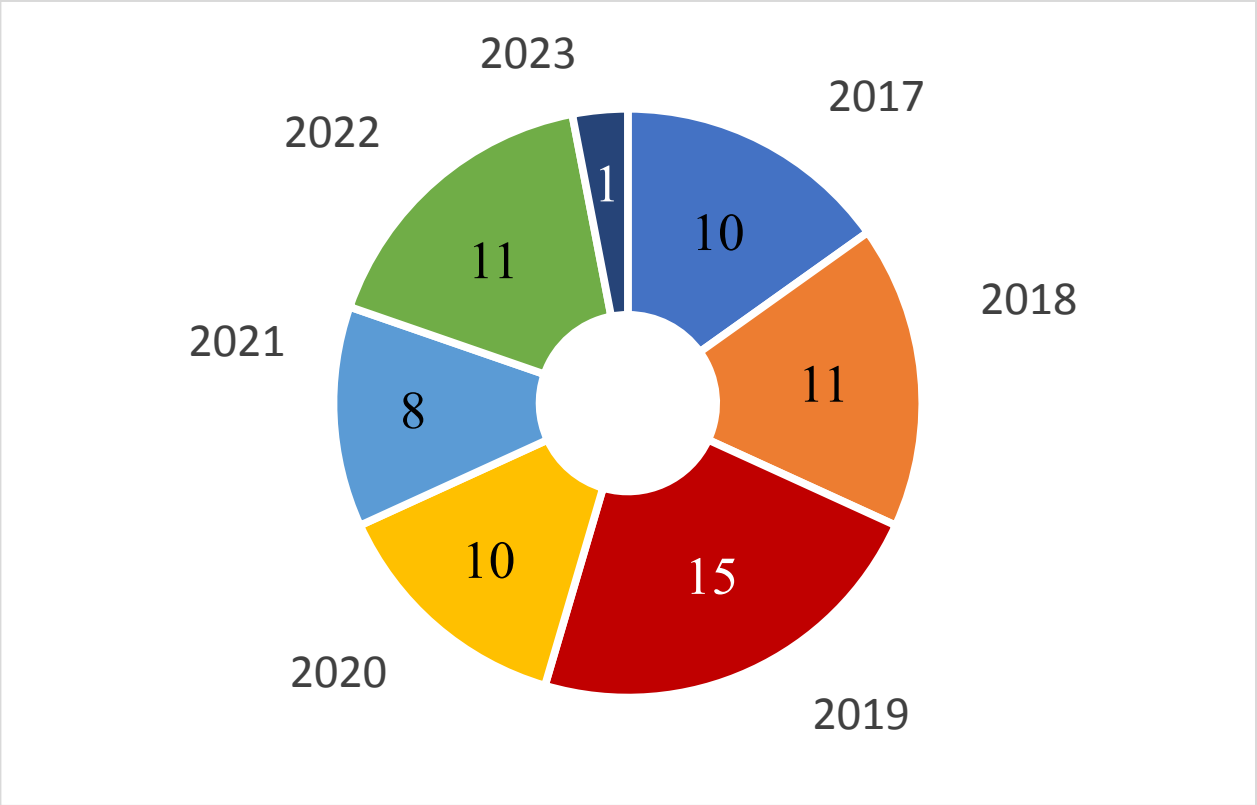

*(a) Year of reviewed articles*

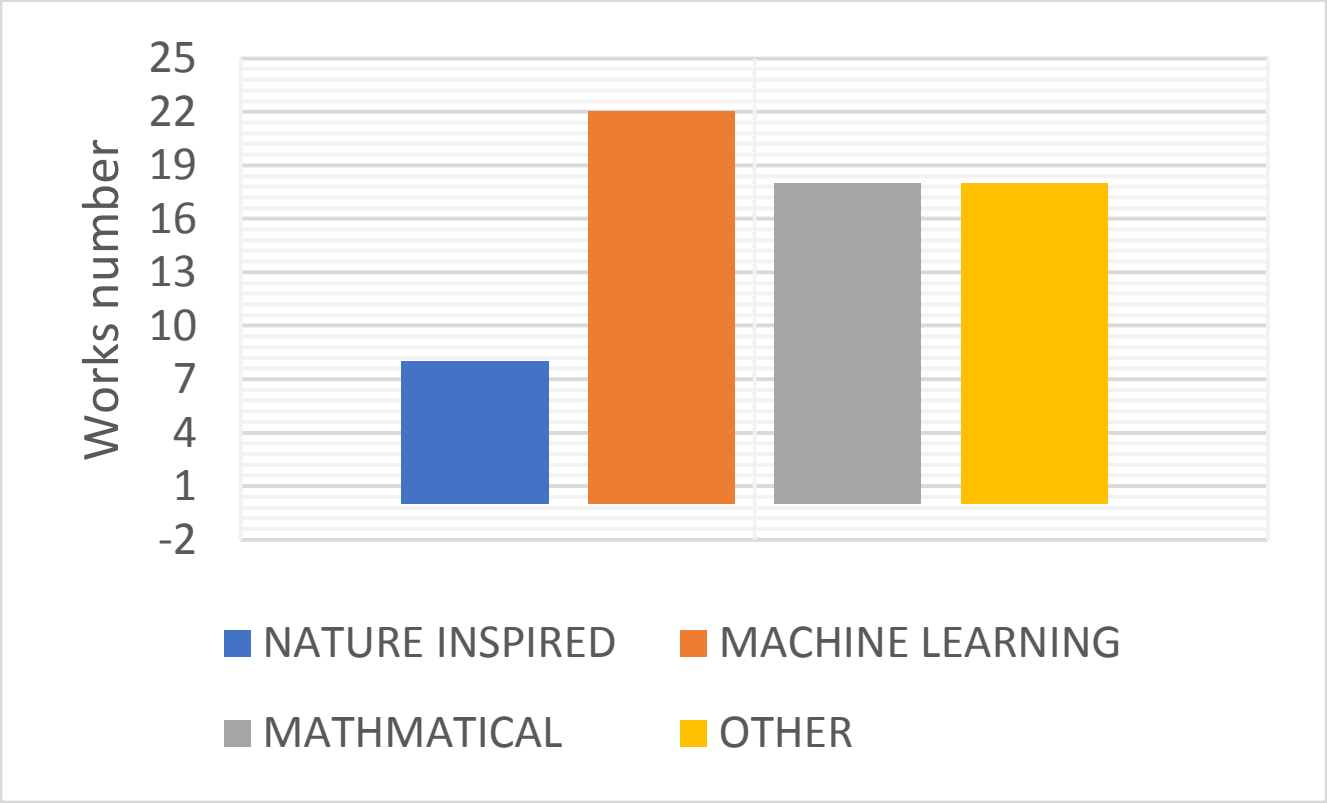

*(b) Reviewed Algorithms categories*

*Figure 3: Reviewed articles publication years and algorithms*

#### 3.1.1. NATURE INSPIRED BASED LOAD BALANCING METHODS

Nature inspired is a term used to describe classes of meta-heuristic algorithms that resemble or are inspired by natural events explained by the scientific sciences [21]. This approach increases the performance of SDN load balancing in terms of reduced overall waiting time, response time, and completion time for resources. The authors in [22] presented a dynamic load balancing solution based on Particle Swarm Optimization (PSO). This study presents an intelligent LB method for controlling resources and running applications on schedule in a cloud environment. In this work, a fitness function was developed to balance loads quickly and efficiently. The authors have asserted that because of their technique, the response time has decreased, throughput has improved, and customer satisfaction has reached the maximum anticipated level. However, the proposed method only effective for applications with relatively limited data. Similarly, in [23], the researchers utilized Type-2 Fuzzy-based Particle Swarm Optimization (TFPSO) to determine the optimal under-loaded local controllers. Also, to predict the future load of local controllers, Markov Chain Model (MCM) is applied. Moreover, a support vector machine (SVM) categorizes the traffic according to its level of importance. The experiment results showed that without priority-based flow classification, this method would overload the network. Research by [24] presented a dynamic approach that uses the Salp Swarm Optimization algorithm (SSOA) and chaotic maps to enhance the optimizer performance. Their technique dynamically establishes the best possible link between switches and controllers and calculates the optimum number of controllers to use. The controller maintains data regarding the global perspective of the whole network. Thus, it allows for the dynamic choice of other routes on demand. In addition, it maintains information for calculating the link's use, checks the latency in the link, and stores load data. However, in the testing process, only some QoS indicators not taken into account.

Furthermore, authors in [25] emphasize hybridizing Bacterium Foraging Algorithm (BFA) and PSO algorithms to enhance QoS multicast routing issue solutions. PSO's ability to transmit social information can be paired with BFA to boost their exploration and exploitation capabilities simultaneously. The proposed approach produces delay-compelled connections to each multicast destination. The bacterium foraging algorithm (BFA) constructs a multicast tree from the minor latency paths collection. To maintain a fair balance between the algorithm's intensification and diversification, the authors dynamically changed PSO's parameters to satisfy global search and BFO, reducing delay and providing an ideal solution. Nonetheless, there is a need to take into account certain supplementary factors such as the mobility and energy limitations associated with mobile devices/sensors, in addition to the Quality of Service (QoS) parameters.

The researchers in[26], [27] combined two AI load-balancing strategies to overcome the SDN LB. In [26], the author examines two strategies, Ant Colony Optimization (ACO) and PSO. In addition, employing a dependable connection and node to design the path to the target node may improve speed and network load balancing. The authors present a conceptual framework for SDN futurology by analyzing node and network resilience to balance the load and improve QoS. Furthermore, the paper presented by [27] used a Genetic algorithm (GA) with ACO to handle load imbalance and convergence lag. In the second step of the search, GA is utilized to decrease the search area, allowing the ACO algorithm to find the trajectories of the LB streams correctly. With the proposed method, the RTT and the packet-delivery rate are significantly enhanced compared to the Round Robin (RR) and ACO algorithms. Similarly, in [28], researchers propose software-defined wireless bacteria-inspired networks created by combining GA and BIN (SDWBIN). They claimed their method could determine the best route for traffic engineering networks that also satisfied the quality-of-service requirements. In addition, their solution provides a reliable QoS architecture that decreases network end-to-end communication delays while simultaneously improving network performance. However, the proposed methods take high processing time, overload the network, and in the case of GA, only a few QoS factors were considered. Furthermore, the ACO-based algorithms need considerable time to update both forward and backward. Furthermore, a study by [29] proposed a method based on the GA to distribute the controllers' loads in SDNs effectively. This strategy used a configurable threshold to recognize the overloaded controller and carefully select the right moment to migrate switches based on different criteria to ensure the best result regarding load balancing. The Jaya algorithm determines the importance of load imbalance and migrations number, which are the criteria this work relies on to select the switch-controller pairs. The result showed an improvement regarding throughput, migrations number and response time compared to other techniques, where these parameters have been improved by 47.25%, 67.98% and 9.38%, respectively. Nevertheless, the work does not considered the energy consumption and more efficient predictive methods can be used to identify the threshold value. Table ٢ shows a comparison of the nature-inspired LB algorithms regarding different aspects including the used algorithm / technique, the problem and the method strength and weaknesses.

*Table 2: Nature inspired based load balancing methods and their properties*

| Authors | Algorithm / Technique | Addressed problem | Strength | Weaknesses |
|---|---|---|---|---|
| [22] | PSO | Dynamic resources and on-demand user application requirements make cloud application load balancing complicated. | • Reduces reaction time<br>• Throughput increases<br>• Utilizes more resources | • Only effective for applications with relatively limited data. |
| [23] | TFPSO | Scalability and load balancing | • Reduced latency<br>• Improved load balancing<br>• Improving throughput | • They used priority-based flow classification<br>• Overloading |
| [24] | SSOA | Multi-controller distributed | • Improved execution time and reliability | • Other QoS indicators not taken into account |
| [25] | PSO with BFO | Multicast routing under multiple constraints | • Delay has been reduced<br>• Reduces cost | • QoS multicast over MANET not taken into account |
| [26] | ACO and PSO | Load balancing | • Reduced latency<br>• Improved load balancing<br>• Reduced loss of packets<br>• Improved round trip time | • Reliability not taken into account<br>• Real-time experiment not considered |
| [27] | GA with ACO | Load balancing | • Reduce data transmission time<br>• Effective research<br>• Reduced loss of packets | • Overloading<br>• Only a few GA factors were taken into account. |
| [28] | GA and BIN | Future-generation network traffic management | • Increased throughput<br>• Effective resources utilization<br>• End-to-end cost reduction<br>• Effective communication performance | • Used for medium-sized network deployments |
| [29] | GA | Load balancing | • Improve throughput<br>• Improve migrations numbers<br>• Improve response time | • Energy consumption not considered<br>• More efficient predictive methods can be used to identify the threshold value. |

### 3.1.2. MACHINE LEARNING BASED LOAD BALANCING METHODS

Several studies have recommended using machine learning (ML) methods in conjunction with the SDN architecture to achieve enhanced routing performance [30]. In the context of Knowledge-Defined Networking (KDN), the article by [31] explains how to provide load balancing by using an Artificial Neural Network (ANN). The KDN uses artificial intelligence to regulate computer networks; its knowledge plane includes comprehensive network analysis and telemetry. The suggested technique, which uses ANN, forecast the network performance based on the latency and traffic metrics across jobs to choose the least-loaded path. In the same area, authors in [32] proposed an SDN-based ANN-LB method. Several parameters have been used by the suggested technique to improve transmission efficiencies, such as overhead, delay, hop count, packet loss, trust, and bandwidth ratio. Based on these parameters, the algorithm balances the network load by analyzing network congestion and choosing the least-laden transmission path. The evaluation results showed an improvement in latency, bandwidth utilization, and packet loss rate. However, in both works the proposed methods need more processing time and resources and works better on a medium-sized network or local optimum.

A Back Propagation Artificial Neural Network (BPANN) has been applied by [33], where it is used to determine the optimal virtual machines (VM) based on factors such as CPU and memory usage and response time. The BPANN is triggered by the controller when a server agent, included in dynamic Agent-Based Load Balancing (DA-LB) architecture, assigns a request to an overloaded VM. The proposed load balancing technique uses SDN's global visibility to transfer VMs in the data center network efficiently. In addition, this technique enhances overall network efficiency and performs well for data transfer, according to the results. The suggested approach optimizes resource usage by increasing processing speed and predicting the loaded VMs in heavy load scenarios. Similarly, a paper by [34] trained a Back Propagation Artificial Neural Networks (BPANN) and K-Mean cluster to predict if a user will access networking equipment seamlessly in the future. The BPNN in use featured three hidden nodes, one output node, and four input nodes. This structure allows the load-balancing method to be implemented under actual service conditions. In this experiment, the authors evaluate the proposed technique's delay times and balancing circumstances with alternative flow forecast algorithms. However, the main disadvantage of this technique, it ignored some services that could impede the process of finding the actual shortest path. On the other hand, the researchers in [35] introduce a novel intelligent SDN-based architecture and a new data transmission optimization technique. The proposed method performs the following tasks; identify the path and required node and predict the traffic flow. The authors applied deep neural networks (DNNs) and Q-learning to identify the optimal route. The experiment results showed that DNNs was the most efficient method for handling the network complex traffic compared to other techniques. Still, essential factors such as scalability, topology changes, loss of packets not take into consideration.

Many research studies have introduced Deep Learning (DL) methods to solve the SDN LB problems. The research by [36] suggests a DL approach for load-balancing SDN-based Data Center Networks (DCNs). The authors rely on the connections' varying load levels to train the DL network. The reaction time of the DL approach for load balancing is compared to that of several ML techniques, including ANN, SVM, and logistic regression (LR). The experimental findings show that the ANN and DL algorithms have faster reaction times than the SVM and LR techniques. Furthermore, DL accuracy is superior to ANN accuracy. The study by [37] described a method as a minimal workload routing algorithm that would choose the network path with the fewest users currently using it. When the likelihood of a system transition information is unknown, the Q learning approach is used to learn and explore that knowledge to provide a near-optimal scheduling node strategy. Similarly, Deep Reinforcement Learning (DRL) is used in this article [38] to properly load balance requests sent to services inside a data center network. Consequently, a strategy capable of dynamically adapting to changing request loads, including changes in the capabilities of the underlying infrastructure. Furthermore, an SDN framework based on machine learning has been proposed by [39]; this framework employs a novel DRL technique known as Deep Deterministic Policy Gradient (DDPG) to improve the routing process in software-defined networks. DROM, DDPG Routing Optimization Mechanism, was proposed to provide real-time, regional, and individual control and administration of network data. The evaluation results showed that this technique is characterized by durability, stability and high productivity, and it has the potential to improve network performance by providing more stable and advanced routing services than currently available solutions. The work by [40] also employs DDPG to select the better path between nodes in a network. The proposed architecture improves DDPG's empirical-playback mechanism's random extraction technique by sampling the experience pool with the SumTree structure. It can increase the convergence rate by extracting a more relevant experience for network updating with more likelihood. Compared to other RL algorithms, the suggested technique improves the SDN throughput with less training time. Nevertheless, the performance of the suggested methods is decreases in case of node failure and it is not treated as a distinct network topology during the experimentation phase.

In addition, [41], [42], and [43] all employ DRL to enhance the quality of service (QoS) metrics of a network. The authors of [41] present a traffic control method close to optimum to optimize the QoS in a hybrid SDN. In addition, an SDN migration sequence is examined to enhance control traffic and improve the optimization results. After that, the DRL method is implemented in the hybrid SDN to solve the problem of split table routing. Finally, the authors test the technique using open-source traffic information. However, both [42], [43] advocated SINET to improve network routing. For optimal network performance, SINET assigns direct control of numerous key routing nodes to a DRL agent that employs dynamic routing strategies. The experiment done on a network of 82 nodes showed that the proposed method lowered network completion time by 32% and was more resistant to topology changes than earlier DRL-based systems.

In the same way, the authors in [44] present a DRL-based technique to generate an SDN route based on human self-learning. This proposal employs deep learning, specifically Bio-Inspired RBM for Bio-Inspired Deep Belief Architecture (BDBA), to find the optimal solution. Basic RBM is included in this bio-inspired approach, as is self-learning based on the limbic system's emotional learning. Every Bio-Inspired RBM uses the reward function R to capture environmental dynamics as network regulations. Otherwise, [45] have provided a Deep Learning Aided load balancing routing approach that combines Queue Utilization with machine learning to control the high and unbalanced load on the router. In order to alleviate the effects of network congestion, they have created a hybrid strategy combining queueing and neural networks and have employed the principal component analysis method to minimize the network's dimensions. In comparison, the study by [46] offers Critical Flow Rerouting-Reinforcement Learning (CFR-RL), a technique based on Reinforcement Learning that automatically develops a strategy to choose critical flows for any given traffic matrix. By creating and solving a basic Linear Programming (LP) problem, CFR-RL reroutes these selected vital flows to balance the network's link use. Still, superior efficiency of using DRL could be only achieved by rerouting a tiny portion of total traffic as the evaluation findings demonstrated. However, authors [47] propose an approach that merges Software Defined Network (SDN) architecture and machine learning technologies. They apply three supervised learning models to categorize data traffic in a software-defined network architecture: Support Vector Machine (SVM), nearest centroid, and Naive Bayes (NB). Then network traffic is studied by capturing traffic traces and creating flow characteristics that pass to the reinforcement learning classifier for prediction.

Alternatively, the authors in [48], [49], [50] determine the degree of load congestion using a Bayesian network and a Long Short-Term Memory (LSTM), respectively. In [48], the authors suggest a load-balancing strategy for IoT controllers that mimics the SDN architecture of conventional data centers. The Bayesian network predicts load congestion by integrating reinforcement learning with self-adjusting parameter weight to balance the load and improve network security and stability. Preemptively balancing the SDN control plane load proposed by [49] facilitates low-latency network flows. Firstly, they anticipate SDN controller demand to prevent imbalances and arrange data plane migrations. Then, the authors optimize migration activities to balance load with delay. In the first step, two prediction models were built using ARIMA and LSTM to forecast the SDN controller's load. The two models were compared regarding the accuracy and predicted mistakes. In the second step, the authors formalized the problem as a nonlinear programming model with binary variables, verified its NP-complete, and suggested a DRL as a solution. Also, research by [50] proposed a dynamic architecture that relies on predicting the link state to balance the load in an SDN efficiently and solve controller-switch transmission delay. The architecture works as follows; the link-state values are predicted using the LSTM algorithm, and then Dijkstra weight is used to find the most efficient route between hosts based on those values. The experimental results showed that the proposed method improves load balancing by 23.7% compared to Open Shortest Path First (OSPF) and 11.7% compared to Q-Learning in the GEANT network. Moreover, it solved controller-switch transmission delay. On the other hand, the authors in [51] devised a mechanism that routes TCP/UDP packet traffic based on numerous factors. They performed K-Means and DBSCAN based on twelve selected factors and determined the appropriate number of clusters to send the request to the appropriate servers. A multiple regression-based searching (MRBS) method has been proposed by [52] to select the best server and path in the data center networks. This method works under high-load situations to enhance network performance. The combination of regression analysis and heuristic algorithm, applied to server statistics information such as load, response time, bandwidth, and server usage, allows MRBS to choose the best server to handle the anticipated traffic. MRBS improves server utilization to 83% compared to traditional algorithms while decreasing delay and response time by over 45%. However, the proposed methods in these works suffer from the following; node migration needs to be considered in case of fault, some QoS parameters must be considered in the evaluation stage, and algorithms must be evaluated in large networks. Table ٣ shows a comparison of machinel learning based LB algorithms regarding different aspects.

*Table 3: Machine learning based load balancing methods and their properties*

| Authors | Algorithm / Technique | Addressed problem | Strength | Weaknesses |
|---|---|---|---|---|
| [31] | ANN | Load balancing | • Improved load balancing | • Used for medium-sized network deployments |
| [32] | ANN | High volumes of traffic which causing unneeded delay | • Improved load balancing<br>• Effective communication performance | • The result is a local optimum |
| [33] | DA-LB | Efficiently use of existing Cloud resources | • Enhances overall network efficiency and performance | • Live migration not supported<br>• Overloading |
| [34] | Neural networks and k-means | The extra latency is caused by load balancer packages. | • Improved load balancing<br>• Effective communication performance | • This technique ignored energy savings. |
| [35] | Q-learning algorithm with Deep Neural Networks (DNNs) | Efficient path selection for batter load balancing | • Load-balancing improvements<br>• optimum in path selection<br>• Predicting flow | • Overhead<br>• Only a few factors were used |
| [36] | DL | Balance the load among servers | • Improved load balancing<br>• Improved response time | • Different typologies not considered<br>• Other QoS indicators not taken into account |
| [37] | Q learning | Multiple controllers load balancing | • Reduced latency<br>• Improving throughput | • Overloading<br>• The result is a local optimum |
| [38] | DRL | Manage various service requests | • Improved load balancing<br>• Reduce host CPU's computational power | • One point failure |
| [39] | DRL | Uniform route optimization | • Network optimization<br>• Reduces delay<br>• Improving throughput | • Other QoS indicators not taken into account |
| [40] | DRL | Traffic engineering throughput issue | • Improve the convergence rate<br>• Better performance and stability | • Topology changes not taken into account<br>• Other QoS indicators not taken into account |
| [41] | Bio-Inspired DRL | Hybrid SDN routing policy | • Improved communication performance<br>• Delay has been reduced | • Handling the scalability problem not taken into account |
| [42] | SINET | Flow routing performance-optimizing | • Improved the robustness and scalability | • Topology changes not taken into account |
| [43] | SINET | Routing optimization | • Reduced flow completion time<br>• Better robustness | • Hierarchical node not considered |
| [44] | DRL | Distributed controller failure | • Improved QoS, security and network policy | • Compared with traditional approach only |
| [45] | Machine learning aided load balance | High and unbalanced load on the router. | • Reduced loss of packets<br>• Improving throughput<br>• Delay has been reduced | • Other QoS indicators not taken into account<br>• Increase loss of packets |
| [46] | CFR-RL | Network disruption impact | • Batter balance link utilization<br>• Improved network | • Delay not taken into account<br>• This technique ignored |
| | | | traffic rerouting | large network |
| [47] | SDN and ML | Network traffic classification | • Improved accuracy of traffic classification | • Lack in real-time traffic collection and categorization of network data |
| [48] | Bayesian network | Load balancing | • Network security and stability improved<br>• Improved load balancing | • Other QoS indicators not taken into account |
| [49] | Preemptively balance with ML | Multi-controller load balancing | • Minimizing the migration time | • Other QoS indicators not taken into account |
| [50] | LSTM & Dijkstra | Load balancing and controller-switch transmission delay | • Improved load balancing<br>• Solve controller-switch transmission delay | • Less improvement when delay is 0<br>• Other QoS indicators not taken into account<br>• Not tested in large-scale networks |
| [51] | Machine learning | Bottleneck | • Improved load balancing | • Other load balancing parameters not considered |
| [52] | Multiple regression | Load balancing | • Improved load balancing<br>• Improved delay and response time<br>• Improved serves utilization | • Not evaluated in cloud computing<br>• Not tested in large-scale networks |

#### 3.1.3. MATHEMATICAL MODEL BASED LOAD BALANCING METHODS

The SDN can be modeled mathematically using algebra, a formal model of transmitting shared resources, or an analytical model employing network calculus [53], [54]. To perform load balancing in software-defined Wi-Fi networks, the authors of [55] have proposed a multi-controller SDN architecture that includes global and local controllers. The global controller utilizes the Analytical Hierarchical Process (AHP) approach to allocate the flow to each controller, where different limitations have been considered based on the local controllers' current state. The global controller is in charge of handling cluster creation and is also responsible for controlling the local controllers, while the local controller is in charge of the local device load, and clustering is regularly updated. Nonetheless, few parameters were used to implement AHP and evalute its performance.

Alternatively, Rounding based Route Joint Deployment (RRJD) algorithm employed by [56], [57], [58], [59], [60]. The authors in [56] focused on hybrid routing as a joint optimization issue and first demonstrated that it was an NP-Hard issue. After that, a RRJD method is applied to fix the issue and boost the network's speed. Likewise, in this research [57], factors such as the control link limitation and other data plane constraints in SDNs were considered to improve QoS. They demonstrate that NP-Hardness exists by explaining the problems of low-latency route deployment and the LB of the control link. In addition, two solutions are presented for each issue with bounded approximation factors and implement the suggested approaches on a tested SDN.

Similarly, A load load-balancing routing mechanism that works on both links and controllers was proposed by [58] (LBR-LC) to solve the NP-Hard overload issue in an SDN. The approach based on rounding has been presented as a solution to the problem since it offers greater scalability and reduces the load. The suggested technique lowers the maximum controller response time by 70% compared to the existing solution but with a 3% increase in the link load. Also, the work by [59] presents a revolutionary SDN-MPLS method with minimal complexity. This method advances bandwidth-restricted routing in mobile networks by balancing network load, route length, energy savings, and network complexity.

Research by [60] examines the issue of how to manage network traffic unpredictability while doing load balancing on commodity switches without the need for extra hardware or software. The article designs and implements the PrePass method, which combines wildcard entries for fine aggregate flows to satisfy the flow table size constraint with reactive routing for newly incoming flows to achieve load balancing despite uncertainties in traffic. The authors present a practical method based on randomized rounding and demonstrate that, in most situations, it may lead to constant bi-criteria approximation. Nonetheless, some Quality of Service (QoS) metrics have neglected, and these techniques required further real-time traffic monitoring and classification of network data.

The fuzzy-logic theory was introduced by [61], [62] to solve SDN LB problems. According to [61], a fuzzy function initially examines the parameters that impact server load and then evaluates the virtual server's load. Based on this, SDN control is employed to keep track of server data throughout the whole network and to implement virtual server tasks. The load and energy usage are dynamically balanced when servers freeze and restart. In the same vein, the authors of [62] evaluated network performance using different measures to create a similar technique that ensured load balancing and enhanced the performance of an SDN. However, the proposed method takes more time to restore the traffic; therefore, some packets will be lost during that period.

A study by [63] introduced a new method for clustering in WSN-based IoT systems, which utilizes Fuzzy C-Means (FCM). The method involves using FCM to create clusters and reducing energy usage in each cluster to determine the optimal Cluster Head (CH). Instead of constantly replacing CHs for dynamic clustering, the study proposes using an energy threshold to determine whether a CH is still functional based on its current energy level, which can extend the lifespan of the sensor network. The suggested FCMDE has the potential to decrease energy consumption and improve durability while keeping expenses low. However, employing metaheuristic optimization methods can improve the CH selection function. In addition, other QoS indicators are not considered during the experiment process. Similarly, the article [64] presented an IoT protocol called EFUCSS, which is an energy-efficient based on fuzzy logic and unequal clustering with sleep schedules, and uses WSN. This protocol aims to increase the network's longevity and decrease energy consumption by employing clustering, scheduling, and data transmission techniques. The proposed protocol used Fuzzy C-Means to create unequal clusters, which reduce the distance data travels and balance energy usage. The cluster heads selection process used a fuzzy logic system that takes input variables such as gateway distance, remaining energy, and centrality. Cluster heads (CHs) collect data from other cluster members, consolidate it, and transmit it to the gateway (GW) in a single hop. A sleep scheduling strategy is employed between the coupled nodes to reduce the number of transmitted nodes. According to the findings, the EFUCSS method could lead to a notable increase in the remaining energy of 26.92% to 213.4% and an extension of network lifespan by 39.58% to 408.13%. Furthermore, EFUCSS is more effective than other comparable algorithms in extending the life of networks. However, the suggested approach for IoT based on WSN does not involve the management of mobile sensor nodes. Also, utilizing mobile sink-based data aggregation and scheduling may enhance the capability of the sensor nodes. Additionally, prediction methods can be utilized to forecast data for nodes that are not currently active. Similarly, A study by [65] proposed an SDN-based architecture to balance traffic across IoT servers and fulfill the QoS requirements for various IoT services. Initially, the authors model the issue as an NP-hard Integer Linear Programming (ILP) instance. After that, they offer a heuristic technique for proactive and predictive QoS management using time-series analysis and fuzzy logic. Finally, the Open vSwitch, Floodlight controller, and Kaa servers are used to build and test the framework. The outcomes showed improvement in IoT QoS metrics like throughput and latency while preventing server overload in high-traffic environments. In terms of performance, the suggested framework beats competing approaches. The suggested framework has better performance than existing techniques. However, the framework has yet to be tested on a distributed SDN control plane or multi-domain network. Moreover, additional factors could improve performance, such as employing progressive policies to estimate the load, implementing Network Functions Virtualization to conserve energy, and enhancing QoS management.

In software-defined elastic optical networks, work by [66] proposed an optimization method to reduce the cost-of-service delay and minimize the load imbalance. The authors proposed a measurement method based on entropy for analyzing load imbalance and developing joint optimization utility functions. The technique works in three stages; firstly, the optimizer selects the possible solutions and then passes them to the defragmentation algorithm for the examination process. Finally, at the end of each connection per wavelength, a power budget algorithm is used to calculate criteria including received power, noise, and OSNR for all network services and use it for validating a routing solution.

Similar work by [67] uses active learning based on entropy to detect intrusion patterns at the packet level efficiently. The suggested approach also can be used to spot assaults on the network. Then, a load-balancing technique applies balancing sensor computing capacity and source requirements to maximize the utility of vehicle sensors. Thus, utilizing a convergence-based technique resulted in maximizing resource consumption.

Bandwidth is one of the elements that must be considered for effective load balancing. An increase in the number of terminals connected to the network will increase the demand for bandwidth and data transport. The authors in [68] proposed an LB architecture to solve the need for more bandwidth and enhance the network performance by using service-oriented SDN-SFC. The method categorized the incoming requests based on their type and assigned priority for each service. A heuristic approach was then used to choose a transmission path from the available service function chains. This method expedited data transfer, and it also enhanced the degree of load balancing. However, using KKT alone to minimize the response time could cause irregular load distribution as it depends on the arithmetic configurations of the controllers. On the other hand, to overcome the controller migration issues, work by [69] presented a new method that utilized the Karush-Kuhn-Tucker (KKT) conditions and the Demand and supply curve-based SDN (DSSDN). KKT is used to solve the response time issue; however, the controllers with fewer computational configurations (fewer routers) will take on fewer burdens. Therefore, the authors employed DSSDN to dynamically select the OpenFlow devices that maximize controller burden while minimizing user traffic. In [70], the authors suggested a non-cooperative load-balancing strategy that builds based on the principles of mean-field game theory. This approach is intended to achieve load equilibrium based on the response time value of each SDN controller. The algorithm makes the routing decision for each request, known as Wardrop equilibrium, which leads to efficient load-balancing. Work by [71] employed dynamic load balancing to optimize resource and bandwidth usage by determining the shortest path to the destination and improving QoS performance. Implementing Dijkstra's method helps locate many pathways of equal length and narrows the search space in the topology. In addition, Priority traffic flows are assigned a specific sequence. It then directs traffic along the route with the lowest cost and load among those considered. Furthermore, the researchers in [72] used SDN's global network view to implement load balancing and reduce network latency by determining the optimum data transmission path. Each route was surveyed for essential elements. Load balancers evaluate features such as throughput, packet loss, latency, hops, and node utilization. These features are the input for a trained neural network that predicts the overall load state for Dijkstra's shortest pathways. But the proposed methods in those works need more accurate traffic prediction in the case of IoT and satellite contexts. Table ٤ shows a comparison of mathematical model-based LB algorithms regarding different aspects.

*Table 4: Mathematical model-based LB algorithms and their properties*

| Authors | Algorithm / Technique | Addressed problem | Strength | Weaknesses |
|---|---|---|---|---|
| [55] | (AHP) | Load balancing the controller with flow request processing | • Improved load balancing<br>• Throughput has increased<br>• Delay has been reduced | • Few parameters were used to implement AHP |
| [56] | RRJD | Flow optimization and load balancing | • Improved load balancing<br>• Simplified flow rules<br>• Faster deployment<br>• Control load is reduced. | • Consideration should be given to the updating strategy and online algorithm. |
| [57] | RRD | Up-links and down-links have a substantial influence on QoS performance because of their limited capacity | • Decrease reaction time<br>• The controller can fine-tune each flow<br>• preventing data and control plane congestion | • Other QoS indicators not taken into account |
| [58] | Rounding-based algorithm | Scalability and load management | • Improved load balancing<br>• Better QoS<br>• Effective controller usage | • Overloading |
| [59] | MPLS-SDN | LSP configuration | • Improved mean blocking ratio<br>• Improved mean CPU time | • Other load balancing parameters not considered |
| [60] | PrePass | SDN switches limited resources load balancing | • Switches resource constraints satisfied | • The link load ratio increased |
| [61] | Fuzzy logic | Server load balancing | • Improved server load balancing<br>• Effective communication performance | • Overhead<br>• Frequent server sleep/restart |
| [62] | Fuzzy logic | Scalability and load balancing | • Improved load balancing<br>• Effective communication performance | • Overloading<br>• Static load |
| [63] | Fuzzy C-Means | The power consumption of sensor nodes | • Enhanced network durability.<br>• Lowered energy consumption<br>• Improved throughput<br>• Reduced latency<br>• Reduced deployment cost | • No optimization procedures used in developing the method for choosing Cluster Heads.<br>• Other QoS indicators not taken into account |
| [64] | EFUCSS | The power consumption of sensor nodes | • Enhanced network durability.<br>• Lowered energy consumption<br>• Extended the network lifespan | • Does not involve the management of mobile sensor nodes.<br>• Utilizing mobile sink-based data aggregation and scheduling may enhance the capability of the sensor nodes<br>• No prediction methods |
| [65] | Fuzzy logic | Load balancing | • Improving the throughput and latency | • Not tested on a distributed SDN<br>• Other QoS indicators not taken into account |
| [66] | Entropy-based | Optimize the distribution of fiber loads throughout the network. | • Balancing fiber load<br>• Reducing service interruption costs | • The result is a local optimum |
| [67] | Entropy-based | Load balancing | • Improved load balancing<br>• Improved response time<br>• Improved energy savings. | • Works better with large size networks |
| [68] | SDN-SFC | Massive demands on the available bandwidth | • Optimal bandwidth usage<br>• Improved load balancing<br>• Reduce data transmission time | • Other load balancing factors including reaction time, job size, and execution time were ignored |
| [69] | KKT | Optimized controller selection for better load balancing | • Improved load balancing<br>• Better QoS<br>• Effective communication performance | • Other AHP indicators not taken into account |
| [70] | Game theory | Load balancing to avoid congestion | • Improved load balancing<br>• Improved scalability<br>• One point failure reduced | • Constancy of convergence qualities<br>• SDN proxies increased |
| [71] | Dijkstra's algorithm | Congestion and load balancing in the network | • Optimal bandwidth usage<br>• Improved load balancing | • Other QoS indicators not taken into account<br>• Overhead |
| [72] | Dijkstra's algorithm | Data transmission delay time | • Improved load balancing<br>• Delay has been reduced | • This technique ignored shortest path discovery |

### 3.1.4. OTHER DYNAMIC LOAD BALANCING METHODS

Researchers in [73] introduced a dynamic SDN LB approach that tackles the timing of service responses as the most critical component for defining the user experience inside the service model. This technique presented a dynamic load balancing mechanism based on server response times (LBBSRT) in SDN architecture, that uses server clusters. The response time of each server is obtained by the SDN controller, which then selects the server with the fastest or most stable response. This approach uses server resources more efficiently than static algorithms such as round robin and random to improve LB performance. The presented approach lowering server response time to balance the networks load. However, the higher the cluster load, the longer it takes to reply, and the more computers in a cluster deliver the same service.

In [74], the researchers indicated that the problem of partial flow statistics collecting (PFSC) is NP-Hard. When balancing loads, it is essential to consider the quality of flow data to reduce the overhead generated by rerouting flows and provide the best possible solution. Therefore, the authors offer an adaptive flow statistics gathering system based on (switch) port statistics information, which enabling effective routing by utilizing link load similarities. In [75], researchers also optimized load balancing and congestion in four stages to record the correct information into the OpenFlow switches' flow tables. In the first two levels, sub-topology reduces space and increases performance. The third phase implements load balancing, while the last phase creates pathways and injects flows in switches. Furthermore, an SDN-Based Load Balancing (SBLB) has been suggested by [76]; this strategy prioritized the SDN to optimize resource utilization and user response time. The proposed mechanism comprises a server pool that communicates with the SDN controller through OpenFlow switch flow tables and an application module that runs atop the controller. Also, a dynamic LB algorithm for data plane traffic presented by [77] mitigates bottlenecks. This method rerouted ties as network usage increases. The method evaluated the connection cost by picking the optimum link after detecting bottlenecks. Then, it decreased network latency and packet loss. The paper claimed that the suggested technique could balance data plane traffic in any SDN setup. Despite its advantages over other approaches, these techniques have a significant flaw: it relies only on reactive flow entry and does not undertake real-time load monitoring.

Through the use of cloud-based SDN, authors in [78] provided a Services Orchestration and Data Aggregation technique (SODA) to address the issues of data redundancy and sluggish service response. SODA can combine data packets to eliminate redundancy and speed up service response. This technique divides the network into data centers, middle routing, and vehicle network layers. Each layer has its task, while the data centers layer is responsible for service response delay reduction through distributing apps with very particular functions to every device on the network. The middle routing layer adjusts the data packets routing path according to the routing distance and the packets' correlation. The last layer, the vehicle network layer, transmits data packets and services between the network equipment. The proposed method did not relied on QoS indicators to prove its efficiency.

The article by [79] suggested a load-balancing algorithm that sees the network as a graph, where the vertices are switches and the edges are the network's channels. They have considered the channel capacity's supremacy over server load while determining the ideal path and proven the approach using a mathematical example. Work by [80] presented SDN dynamic offloading service for an SDN-based fog computing system to choose the optimum offloading node and assist the offloading path. The proposed system selects the best offloading node based on current computational characteristics and network resource information. The results showed outperforming existing strategies that do not employ SDN technology in terms of throughput, request-response time, and end-to-end bandwidth guarantee. The study by [81] provided a multi-path routing system based on SDN that employs several characteristics, such as latency, bandwidth, and node load. By detecting the network state and switching node load, the method builds a model to compute link transmission cost to adjust the end-to-end transmission path in real-time. This work used Systems Tool Kit (STK) to create the inter-satellite propagation delay model to enhance the estimation rate of the transmission cost. The lack of QoS factors, such as cost optimization and network latency, is a significant flaw in the suggested approachs.

Load imbalance among statically configured controllers is a critical issue. Researchers in [82] presented the Assessing Profit of Prediction (APOP) method to overcome these issues. This technique balances the network load by evaluating the profit and predicting the overloaded. They offer Taylor's method to anticipate network flow change and analyze the profit of moving switches in advance to save migration time and avoid detrimental consequences. Also, the paper by [83] proposed the online controller load balancing (OCLB) approach, which focuses on balancing the load by minimizing the controller's average reaction time. This method executed switch migration sequences with various real-time applications and a consistent parameter in mind, all with the end goal of minimizing response time. The results showed that the proposed scheme executed online to provide almost optimum load balancing over the control plane. Similarly, to solve the migration problem, the authors in [84] integrated swap movements and shifts into a search system. In contrast to the methods already in use, the suggested algorithm will not quit the search if a switch migration cannot work. Rather than relying on basic techniques, it searches for more advanced operations to enhance velocity, such as switching between two distinct keys.

In addition, the authors in [85] used a multi-controller to deal with traffic loads caused by several nodes. Using a method called load optimization and anomaly detection (LOAD), they were able to reduce the cost of migration while simultaneously increasing controller performance and decreasing reaction times. In contrast, two switch migration strategies, a balanced controller (BalCon) and BalConPlus were introduced in [86], which both boost the capability of regulating traffic load variations. The first is utilized if serial processing of switch requests is not required; otherwise, BalConPlus is used under rest situations, according to them. The researchers claimed that their method considerably minimizes load polarity among the controllers in the network.

Another view is by [87], where the authors analyzed SDN controller load and handover latency, showing that over-loading can lengthen handover latency and use load balancing to prevent it. Their LB management approach used network heterogeneity and context-aware vertical mobility. It has three parts; first, the users are chosen based on context, then the load distribution is minimized amongst controllers, lowering processing and communication overhead. After determining potential users, the algorithm optimizes the selection of diverse candidate networks. These approaches can accomplish the optimal solution, but the runtime and latency in the large-scale network will increase. Moreover, this technique's primary issue is loading fluctuation, which could occur if the target controller and migration switch are incorrectly identified.

The article by [88] suggested a technique for wireless sensor networks termed perceptually important points-based data aggregation (PIP-DA). This strategy aims to reduce the quantity of data readings transmitted, resulting in less energy consumption and a longer lifespan for the network. In addition, the PIP-DA maintained the precision of the data readings collected at the base station. A cluster topology is used to develop the aggregating data within the PIP. The assumption is that the topology already exists, so the suggested method is usable for any clusters formed by clustering protocols. The primary goal of this technique is to decrease the amount of sensed data at the sensor node level to extend the lifespan of the WSN. The proposed method outperforms previous techniques regarding data remaining after aggregation, sets transmitted to the CH (Cluster Head), data correctness at CH, and energy usage. However, an additional dynamic segmentation algorithm can be utilized at both the sensor node and gateway levels to predict the missing data at the CH and improve proposed method. Also, the authors of [89] introduced a method called "DAEP" for conserving energy in WSNs by aggregating data performed at the individual sensor node level based on extracting extrema points. The proposed technique aims to reduce energy consumption and prolong the lifespan of the WSN. The suggested approach operates periodically and comprises three stages in each cycle: data collection, aggregation, and transmission. The method's effectiveness was evaluated by running several simulations using actual sensor data gathered at Intel Berkeley labs and comparing the results with previous research. The findings indicate that DAEP can significantly reduce the amount of data transmitted by 69-80% and energy consumed by 73-77% while maintaining reasonable accuracy levels, making it a promising approach for reducing the load on sensor nodes. Unfortunately, AI, ML, and statistical techniques were not utilized, which could extend the WSN lifespan. Also, the method was not tested on a real sensor network. Finally, a centralized network clustering strategy used by [90] in which the base station (BS) splits the network into clusters and identifies the node with the most significant energy level as the cluster head (CH) for each cluster at the start of the protocol. It picks and rotates CHs within clusters depending on node energy levels before transferring data to the BS to decrease energy usage. This study employed a stationary algorithm with fixed clusters. The number of CHs in the network depends on the network's topology. Compared to the MOFCA and IGHND, the proposed ESCA approach effectively addresses the energy consumption problem and significantly increases the network's lifespan. However, to evaluate the proposed strategy, it should consider the mobility of nodes and obstacles in the area of interest. Table ٥ shows a comparison of the other dynamic LB algorithms regarding different aspects.

*Table 5: Other dynamic LB algorithms and their properties*

| Authors | Algorithm / Technique | Addressed problem | Strength | Weaknesses |
|---|---|---|---|---|
| [73] | LBBSRT | Hardware limits load balancing to server response times. | • Reduces cost<br>• High reliability<br>• Rich extensibility | • This technique ignored server LB energy savings |
| [74] | PFSC | Overhead load balancing via flow rerouting. | • Load-balancing improvements<br>• Overhead reduced | • Normalized traffic flow<br>• Limited controllers |
| [75] | Flow statistics | Network congestion, load balancing | • Increased throughput<br>• Improved load balancing<br>• Reduces reaction time | • Lacks dependability, scalability, and network performance measurement |
| [76] | Statistical information | Increase resource usage and reduce user response time | • Optimal server utilization<br>• Reduces average reaction time | • No graphical user interface<br>• performance evaluation of the suggested load balancing method ignored |
| [77] | Dynamic load balancing algorithm | Distribution of data plane traffic | • Decreases network latency and packet loss | • Various network topologies in a clustered environment not taken into account |
| [78] | SODA | Redundancy data and service response time | • Reduce data redundancy<br>• Reduces average reaction time | • Other QoS indicators not taken into account |
| [79] | Feature extraction | Low channel capacity causes congestion and decreases the reliability | • Improved reliability<br>• Improved network performance | • Other QoS indicators not taken into account |
| [80] | Feature extraction | Selecting the offloading node to handle an overloaded | • Improved request and response time | • Other QoS indicators not taken into account |
| [81] | Feature extraction | Network congestion and delay | • Improved throughput<br>• Optimal bandwidth usage | • Other QoS indicators not taken into account |
| [82] | APOP | Load imbalance among multiple controllers | • Reduces load balancing migration costs<br>• Delay has been reduced | • Other QoS indicators not taken into account |
| [83] | OCLB | Scalability and load management | • switches migration minimized<br>• Improved load balancing<br>• Reduces reaction time | • Other QoS indicators not taken into account<br>• Diverse flow requests ignored |
| [84] | Heuristic approach | Switch migration problem (SMP) | • Improved load balancing<br>• Choosing the best possible controllers | • Migration cost not taken into account<br>• One point failure<br>• Lack of bandwidth use |
| [85] | Switch migration | Load traffic handling | • Reduces run-time<br>• improving the migration cost<br>• Improved execution time | • LOADS scheme can lead to load imbalance due to failure of any actively distributed controllers |
| [86] | Switch migration | Network traffic classification | • Improved accuracy of traffic classification | • Lack in real-time traffic collection and categorization of network data |
| [87] | Heterogeneous networks | Controller response time | • Improved load balancing<br>• Reduces average reaction time | • Minimum channel bandwidth not guaranteed |
| [88] | Perceptually Important Points Based | The power consumption of sensor nodes | • Decreases the amount of extra work on the sensor | • Introducing an additional dynamic segmentation |
| | Data Aggregation (PIP-DA) | | node level.<br>• Lowered energy consumption up to 93% | algorithm that can be utilized at both the sensor node and gateway levels.<br>• Other QoS indicators not taken into account |
| [89] | Data aggregation based on the extraction of extrema points (DAEP) | Processing the data efficiently while conserving energy | • Reduce the amount of data transmitted by 69-80%<br>• Lowered energy consumption by 73-77% | • Artificial intelligence, machine learning, and statistical techniques could be utilized to prolong WS.<br>• Not applied to a real sensor network |
| [90] | Energy-Saving Clustering Algorithm (ESCA) | Network energy consumption and its lifespan | • Increased the network lifetime from 62% to 85% and its expansion from 19% to 53%.<br>• Lowered the network energy consumption | • Mobility of nodes not taking into account.<br>• Other QoS indicators not taken into account |

According to the reviewed articles, critical challenges regarding SDN load balancing have yet to be examined exhaustively and thoroughly. One of the significant issues was the failure of a centralized controller, which could lead to the collapse of the entire network. The suggested solution was distributing the controllers into several domains, but the controller deployment cost will increase, and the controller efficiency to handle network change will decrease. Most of the research examined has yet to demonstrate the impact of the load balancing mechanism on all Quality of Service (QoS) parameters. Also, most of the techniques studied need to consider the challenge of conserving energy and reducing carbon emissions. Incorporating these factors could enhance the effectiveness and popularity of current load-balancing mechanisms. Finally, more research must be conducted incorporating artificial intelligence techniques on load balancing. A hybrid approach that effectively integrates two or more methods can be used to balance the load and improve network performance efficiently.

### 3.2. SDN LB ALGORITHMS EVALUATION METRICS

During the process of evaluating LB algorithms, a set of metrics must be taken into account to prove the effectiveness of those algorithms. The researchers use a wide range of metrics to outline the benefits and drawbacks of the available approaches. This section describes the most used parameters mentioned in the selected papers.

- Response Time (RT): The response time is an essential parameter for LB methods. It is the time it takes for a user to get the info they requested after submitting a query. It is affected by different variables, including bandwidth, network users, requests, and processing time. It is calculated using Equation (1), where $t_1$ is the request submission time, and $t_2$ represents the request start processing time.

$$RT = \Delta(t_1 - t_2) \qquad (1)$$

  Handling a large number of requests in a short amount of time can improve the response time.
- Throughput (T): Refers to the proportion of job requests that were scheduled within a specific time frame (t) and were successfully executed and processed, compared to the total number of completed job requests. High throughput is required for the load balancing mechanism to function correctly. It is calculated using Equation (2).

$$T = \sum_{requests\ i}^{n} (time\ t) \qquad (2)$$

- Resource Utilization (RU): Represents the efficient network's resource utilization ratio during the request processing (e.g., memory, CPU, etc.). It is essential in the LB evaluation process; high RU means the LB algorithm performs well. It is calculated using Equation (3), where $ET$ is the execution time.

$$RU = \frac{\sum_{request\ i}^{n} ET}{Maxrequest\ ET} \quad (3)$$

- Latency: The latency measures how long it takes a packet of data to move across a network. It considers both the delay during transmission and propagation resulting from the packet forwarding process. It is calculated using Equation (4), where $L$ is the latency, $S_{td}$ and $D_{td}$ are the source and destination transmission delay, respectively, $S_d$ represents the switch delay, and $P_d$ represents the propagation delay.

$$L = S_{td} + D_{td} + S_d + P_d \quad (4)$$

- Work load degree: This metric is used to assess the load distribution throughout the networking components. It can be calculated by using a variety of indices, including Jain's fairness index and load balance rate.
- Deployment cost: The cost of network elements deployment includes CAPEX and OPEX. This metric is essential to minimize the SDN implementation cost by calculating the best number of the needed controllers to build an efficient network.
- Jitter: It is the term used to describe the variation in packet transmission time between networking elements; when there is network congestion, the jitter will increase.
- Packet loss ratio: It is calculated by subtracting the number of transmitted and received packets between the source and destination. It occurs when at least one informational package fails to accomplish its goal. LB algorithms always aim to have a low packet loss rate to guarantee efficiency.
- Delay: Represents the time a packet takes to get from one node to another; it includes communication, routing, processing and migration delay.
- Round trip time (RTT): The time it takes a packet to travel from its source to its destination and back again is called its round-trip time. It's a key performance metric for evaluating the efficiency of LB methods. The timeouts will be ineffective since they will be longer than necessary if the round journey takes less time than expected. It is calculated using Equation (5), where $AV_{RTTs}$ is the average round-trip time in the server and $AV_{RTTc}$ represent the average round-trip time in the client.

$$RTT = AV_{RTTs} + AV_{RTTc} \quad (5)$$

- Bandwidth utilization ratio (BU ratio): This metric checks the load placed on the links by assessing the network's transmission capabilities. The link's bandwidth ratio calculates by the SDN controller based on the total number of bytes sent at the associated switch ports during two consecutive periods.
- Migration delay: It is the amount of time that must elapse from when a packet is moved from one switch to another until it reaches its final destination. The number of migrations should be kept to a minimum for effective communication.
- Link utilization: It represents packet transmission speed throughout the communication between the networking components and includes the uplink/downlink rate. It is calculated using Equation (6), where $Lu_{ij}$ represent the link utilization value between two nodes $i$ and $j$. While $b_{ij}$ indicates the link bandwidth and $u_{ij}^{t}$ is the amount of bandwidth utilized during the time frame t between the $i$ and $j$ nodes.

$$Lu = \left[ (Lu)_{ij} = \frac{u_{ij}^{t}}{b_{ij}} \right]_{N*N} \quad (6)$$

- Flow completion time (FCT): It is used to determine the efficiency of flow transportation in data center networks. It represents the amount of time required to finish transferring a file within a flow. LB methods aim to keep the flow completion time as short as feasible.
- Migration cost: Two preliminary charges are involved: the load cost and the cost of sending messages. During the switch migration process, messages such as migration, role requests, and asynchronous messages are needed to transmit between the controllers.
- Overhead: It represents the total sum of all the extra time, space, data transfer, and processing power that the activity requires. Overhead includes communication, flow stealing, synchronization and flow statistics collection overhead.
- Packet load ratio (PLR): This metric is introduced to measure the route performance and calculate the maximum traffic load on each link.
- Power consumption: Represent the amount of energy each node in the network uses to process a request, whether that request is successful or not. Effective LB reduces power use.
- Consumer Satisfaction (CS): It is the overall customer attitude or behavior addressing the discrepancy between what customers expect and what they receive.
- Cumulative distribution function (CDF): This metric is used to determine whether network links are congested by checking if the required flow entries exceed the flow table size on all switches to avoid flows dropping.

This part presents the metrics that most authors consider when conducting state-of-the-art research. The LB algorithm design and development process depend primarily on these metrics, and it is used to assess the algorithm's performance in SDN-based applications. Some of these parameters are widely used by researchers, such as response time, throughput, RU, latency, delay, workload degree, deployment cost, packet loss ratio, link utilization, and overhead. However, many QoS metrics have yet to be considered in the algorithm evaluation process; Table ٦ presents the metrics used to measure the proposed LB approach performance.

*Table 6: Metrics employed in SDN AI-Based load balancing techniques*

| Authors | Response time | Throughput | RU | Latency | Work load degree | Deployment cost | Jitter | Packet loss ratio | Delay | RTT | BU ratio | Migration Delay | Link utilization | FCT | Migration cost | Overhead | PLR | power consumption | CS | CDF |
|---|---|---|---|---|---|---|---|---|---|---|---|---|---|---|---|---|---|---|---|---|
| [22] | ✓ | ✓ | ✓ | | | | | | | | | | | | | | | | ✓ | |
| [23] | | ✓ | | ✓ | ✓ | | | | | | | | | | | | | | | |
| [24] | | | | ✓ | | ✓ | | | | | | | | | | | | | | |
| [25] | | | | ✓ | | ✓ | ✓ | | | | | | | | | | | | | |
| [26] | ✓ | ✓ | | ✓ | | | | | | | | | | | | | | | | |
| [27] | | | | | | | | ✓ | | ✓ | | | | | | | | | | |
| [28] | | | | | | ✓ | | | | | | | | | | | | | | |
| [29] | ✓ | ✓ | | | | | | | | | | ✓ | | | | | | | | |
| [31] | | | ✓ | | | | | | | | | | | | | ✓ | | | | |
| [32] | | | | ✓ | | | | ✓ | | | ✓ | | | | | | | | | |
| [33] | | | | ✓ | | | | | | | | ✓ | | | | | | | | |
| [34] | | | | ✓ | ✓ | | | | | | | | | | | | | | | |
| [35] | | | | ✓ | | | | ✓ | | | | | | | | | | | | |
| [36] | ✓ | | | | | | | | | | | | | | | | | | | |
| [37] | | ✓ | | | ✓ | | | ✓ | | | | | | | | | | ✓ | | |
| [38] | | ✓ | | | ✓ | | | ✓ | | ✓ | | | | | | | | | | |
| [39] | | ✓ | | ✓ | | | | | | | | | | | | | | | | |
| [40] | | ✓ | | | | | | | | | | | | | | | | | | |
| [41] | | | | | | | ✓ | | ✓ | | | | ✓ | | | | | | | |
| [42] | | | | | | | | | | | | | | ✓ | | | | | | |
| [43] | | | | | | | | | | | | | | ✓ | | | | | | |
| [44] | | ✓ | | | ✓ | | | | ✓ | | | | | | | | | | | |
| [45] | | ✓ | | | | | | ✓ | ✓ | | | | | | | | | | | |
| [46] | | | ✓ | | ✓ | | | | ✓ | | | | ✓ | | | | | | | |
| [47] | | | ✓ | | | | | | | | | | | | | | | | | |
| [48] | | | | | ✓ | | | | | | | ✓ | | | | | | | | |
| [49] | | | | | ✓ | | | | | | | | | | ✓ | | | | | |
| [50] | | ✓ | | | ✓ | | | | ✓ | | | | | | | | | | | |
| [51] | | | | | ✓ | | | | | | | | | | | | | | | |
| [52] | ✓ | | | | ✓ | | | | ✓ | | ✓ | | | | | | | | | |
| [55] | | ✓ | | ✓ | | | | | | | | | | | | | | | | |
| [56] | | ✓ | | | | | | | | | | | ✓ | | | ✓ | | | | |
| [57] | | | | | ✓ | | | ✓ | | | | | ✓ | | | | | | | |
| [58] | ✓ | | | | ✓ | | | | | | | | | | | | | | | |
| [59] | | | ✓ | | | | | ✓ | | | | | ✓ | ✓ | | | | | | |
| [60] | | ✓ | | | | | | | | | | | ✓ | ✓ | | | | | | ✓ |
| [61] | ✓ | | ✓ | | | | | | | | | | | | | | | | | |
| [62] | ✓ | ✓ | | | | | | | | | | | | | | | | | | |
| [63] | | ✓ | | ✓ | | ✓ | | | | | | | | | | ✓ | | | | |
| [64] | | | | | ✓ | | | | | | | | | | | ✓ | | | | |
| [65] | | ✓ | | | ✓ | | | | | | | | | | | ✓ | | | | |
| [66] | | ✓ | | | ✓ | | | | | | | | | | | | | | | |
| [67] | | ✓ | ✓ | | | | | | | | | | | | | | | | | |
| [68] | | | | | ✓ | | | | | ✓ | | | | | | | | | | |
| [69] | | ✓ | | ✓ | | | ✓ | ✓ | | | | | | | | | | | | |
| [70] | | ✓ | | | | | | | ✓ | | | | | | | | | | | |
| [71] | | | ✓ | ✓ | | | | | | | | | | | | | | | | |
| [72] | | ✓ | | ✓ | | | | | | | | | | | | ✓ | | | | |
| [73] | ✓ | | ✓ | | | | | | | | | | | | | | | | | |
| [74] | | | | | | | | ✓ | | | | | | | | | ✓ | | | |
| [75] | ✓ | ✓ | | | | | | | | | | | | | | | | | | |
| [76] | ✓ | ✓ | | | | | | | | | | | | | | | | | | |
| [77] | | ✓ | | | | | | ✓ | | | | | | | | | | | | |
| [78] | ✓ | | | ✓ | | | | | | ✓ | | | | | | | | | | |
| [79] | ✓ | | | | | | | ✓ | ✓ | | | | | | | ✓ | | | | |
| [80] | ✓ | ✓ | ✓ | | | | | | | | | | | | | ✓ | | | | |
| [81] | | | | | | | | | ✓ | | ✓ | | | | | ✓ | | | | |
| [82] | ✓ | | | | ✓ | | | | | | | ✓ | | | | | | | | |
| [83] | ✓ | | | | ✓ | | | | | | | ✓ | | | | | | | | |
| [84] | | ✓ | | | ✓ | | | | | | | ✓ | | | | | | | | |
| [85] | ✓ | ✓ | | | | | | | | | | ✓ | | | ✓ | | | | | |
| [86] | ✓ | | | | ✓ | | | | | | | | | | ✓ | | | | | |
| [87] | | ✓ | | | | | ✓ | | | | | | | | | | | | | |
| [88] | | | | | | | | ✓ | | | | | | | | | ✓ | ✓ | | |
| [89] | | | | | | | | ✓ | | | | | | | | | ✓ | ✓ | | |
| [90] | | | | | ✓ | | | | | | | | | | | | | ✓ | | |

As seen in figure 4, response time, throughput, latency, and workload degree are essential metrics considered by many works to evaluate the proposed LB algorithm.

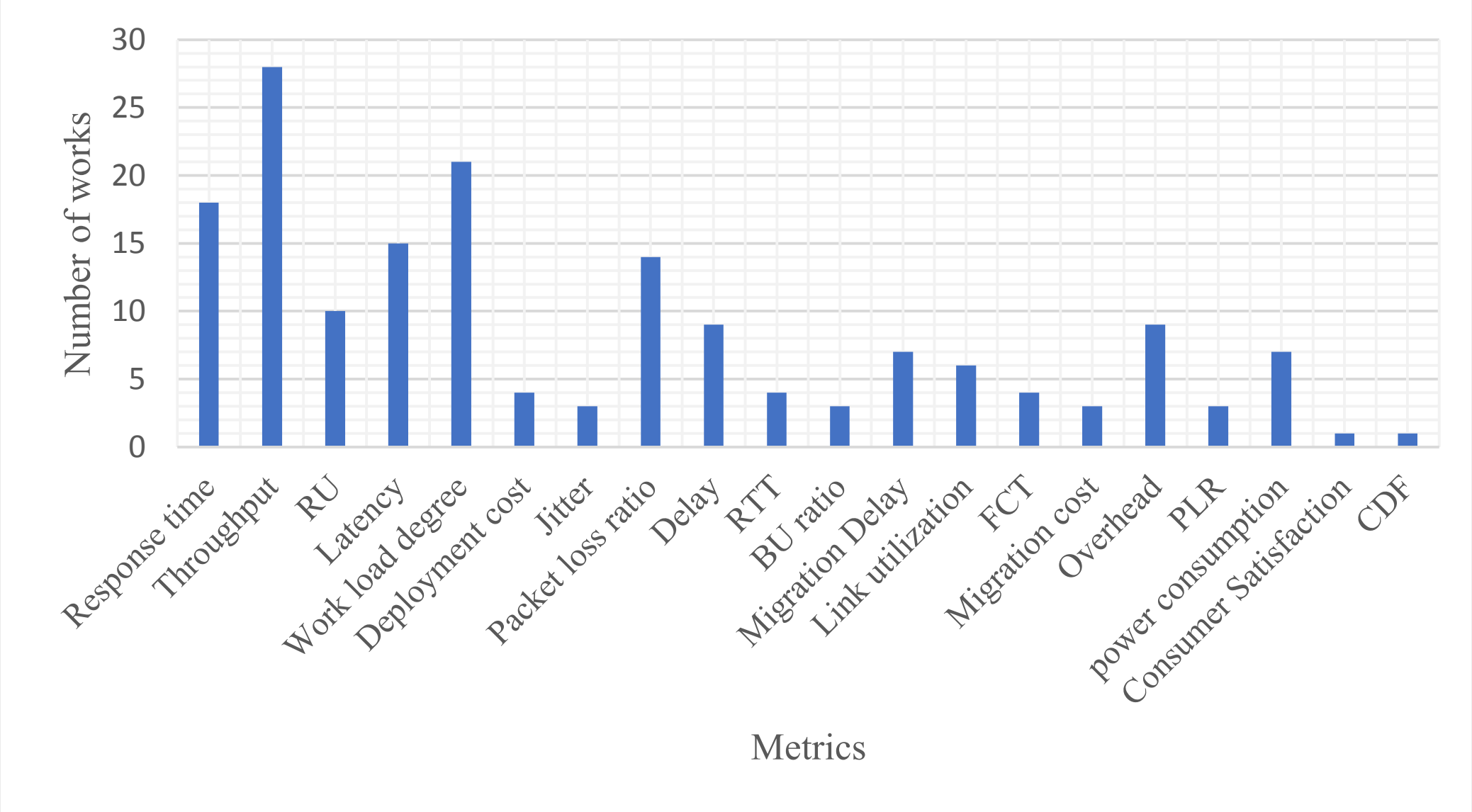

*Figure 4: Evaluation metrics of reviewed articles*

## 4. RESULTS AND DISCUSSION

In this section, we will summarize and compare the different metrics applied by several LB methods published in the past few years. Moreover, we investigated LB strategies that can be used to balance the load in an SDN efficiently and analyzed the limitation of each technique to support innovation in the SDN research field. We evaluated recent studies from reputable journals and conferences to find the most frequent LB performance-enhancing tactics with a particular emphasis on AI-based LB. Tables 1, 2, 3, and 4 compare the four categories regarding different aspects such as the algorithm/technique, the addressed problem, strengths and weaknesses.

This study focuses on AI-based LB techniques and divides them into nature-inspired, machine learning, mathematical model and other LB methods. These mechanisms use LB algorithms that manage work distribution based on each node's actual load and output and adjust the load at the proper time to ensure that the network operates effectively and smoothly. Nevertheless, these techniques have various drawbacks, including the need to respond to burst traffic, dynamically alter the load of the controllers, and neglect some services.

Algorithms based on PSO [22], [23], [24], [25] offered a dynamic LB solution that applies to various service types to ensure the high performance of SDN and customers satisfaction. However, several limitations persist, such as; resource overloading, priority-based flow classification, effective only with relatively limited data and many QoS indicators are not taken into account. On the other hand, heuristic algorithms [26], [27], [28], such as GA and ACO, which are combined with POS and BIN, all of them operate relatively better for huge SDN. In addition, [29] utilized GA to balance the load in distributed SDNs. Still, they take high processing time, overload the network, and in the case of GA, only a few QoS factors were considered. Also, the ACO-based algorithms need considerable time to update both forward and backward.

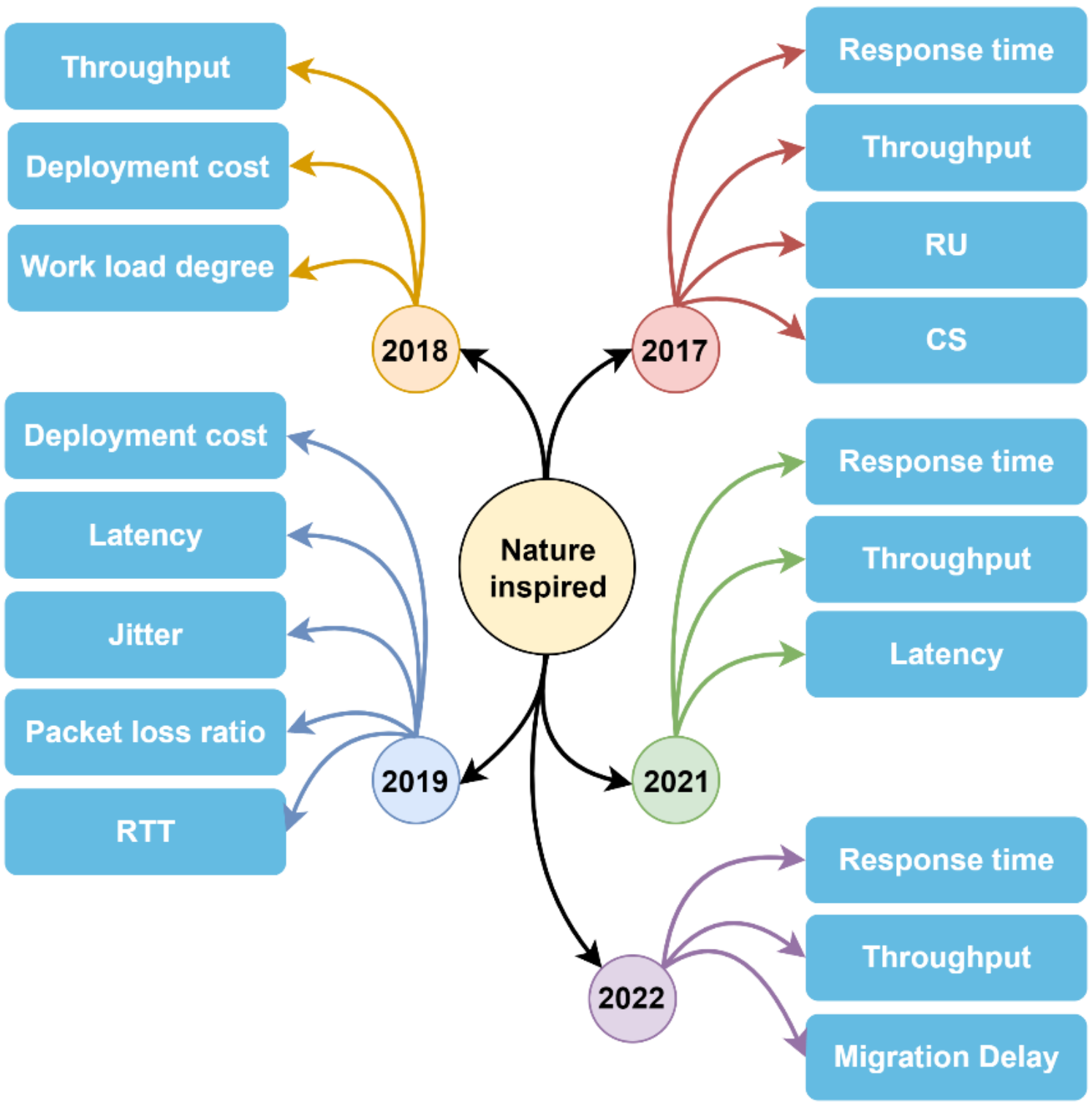

*Figure 5: The metrics trends used by nature-inspired-based LB*

Figure 5 presents the metrics used to evaluate the nature-inspired-based LB in the surveyed papers. Most of the works used response time, throughput, and latency to judge their method's efficiency. Also, some of the studies that conducted from 2017 to 2022 used other parameters in the evaluation process, such as deployment cost, work load degree, jitter, packet loss ratio, RTT, RU, CS, and migration delay. However, many QoS indicators need to be taken into account, where papers included in this survey used only one to four factors to prove the efficiency of the proposed algorithms.

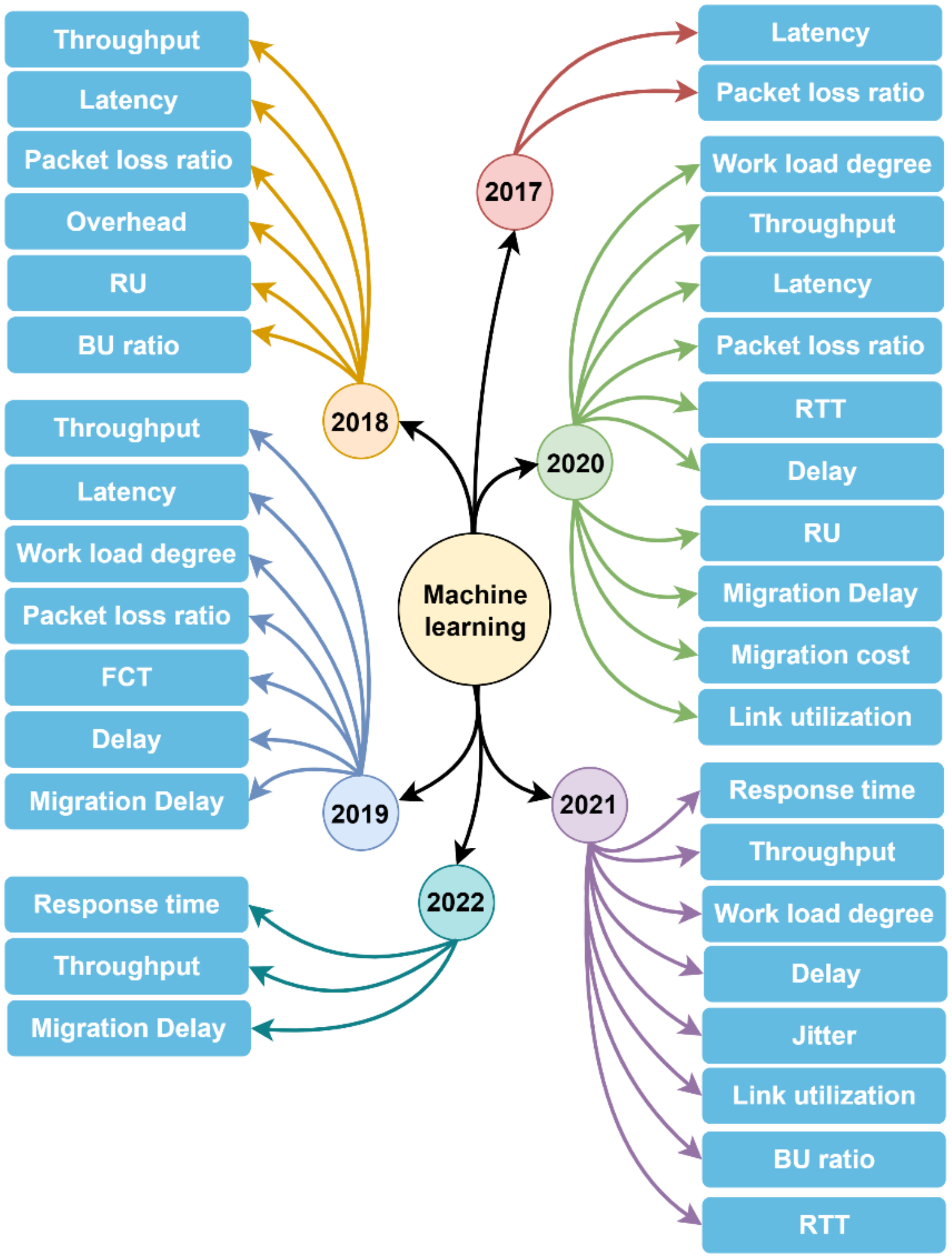

*Figure 6: The metrics trends used by ML based LB*

Using machine learning (ML) in conjunction with the SDN architecture has proven efficient in achieving enhanced routing performance. Based on the graphical results presented in Figure 6, the proposed LB methods that adopt ML have used various metrics in the evaluation process. The majority of the papers that conducted from 2017 to 2022 have used throughput, workload degree, and latency for efficiency proven. However, parameters such as packet loss ratio, delay, migration delay, and migration cost have been used in a few works. Also, publications included in this study employed just one to four parameters to demonstrate the effectiveness of the suggested algorithms.

ANN has been proposed as a solution for load balancing in SDN [31], [32]; this technique improved transmission efficiency using packet overhead, delay, hop count, packet loss, trust, and bandwidth ratio. However, it needs more processing time and resources and works better on a medium-sized network or local optimum.

BPNN algorithm has been applied in [33], [34] to help predict the best path with the most negligible load. The results showed a general improvement in network performance, especially concerning network latency. However, the primary disadvantage of this technique, it ignored some services that could impede the process of finding the actual shortest path. DNN is also used by [35] to choose the optimal route as well it is appropriate for handling the network traffic. Based on the DNN results, the abnormal flow is prevented by predicting the flow rules and identifying all the significant nodes. However, essential factors such as scalability, topology changes, loss of packets not take into consideration.

A deep learning technique is employed by [36], [37], [38], [39], [40] to develop a mapping link between states and behaviors to expedite the problem-solving process of determining the best approach for all conditions. The experiment shows that DL algorithms improve both the LB and response time. As a result, these strategies can dynamically adapt to shifting request loads, including adjustments to the capabilities of the underlying infrastructure. However, it suffers from poor functionality when node failure happens and it is not considered a different network topology in the experiment process.

A reinforcement learning scheme has been introduced by [41], [42], [43], [44], [45], [46] to solve the rerouting optimization problem. This scheme uses a heuristic approach to automatically learn essential flow selection strategies without being given any guidelines for specific rules. The evaluation's findings demonstrated that RL might achieve superior efficiency by rerouting a tiny portion of total traffic.

Other ML algorithms, such as SVM, Bayesian network, LSTM, K-Means, DBSCAN and multiple regression, are applied by [47], [48], [49], [50], [51], [52] to solve the LB issues. All algorithms showed promising results in reducing response time. However, node migration is not considered in case of fault, some of the QoS parameters are not taken in consideration in the evaluation stage and algorithms not evaluated in large networks.

Many studies have used mathematical models to solve SDN load-balancing problems and choose the most efficient decisions. AHP is used by [55] to compare several routes and determine the optimal one regarding different limitations. This technique could enhance the decision by using consistency measures. The method organized the controllers into global and local; they are responsible for handling cluster formation and local device load, respectively. While the approach improved load balancing, increased throughput and reduced delay, using other variables during the implementation process may lead to the best results. Rounding based approach is another mathematical model presented by [56], [57], [58], [59], [60] as a solution to the problem since it offers greater scalability and reduces the load. The method improved load balancing, reduced response time, and prevented data and control plane congestion. However, other QoS indicators are not considered, and it needs more real-time traffic collection and categorization of network data.

Another branch of artificial intelligence known as fuzzy logic is employed by [61], [62], [63], [64],[65] to address the load balancing issue in SDN. In this approach, the flow-handling rules at the controller are used to dynamically calculate and adjust the paths depending on the network's global perspective. The experiments showed that this mechanism efficiently detects faulty links instantly and chooses a backup path. However, it takes more time to restore the traffic; therefore, some packets will be lost during that period. Also, the entropy-based method proposed by [66], [67] to minimize load imbalance and reduce the cost of the required service. This mechanism identifies network bottlenecks and installs new links where needed. However, it does not consider other QoS parameters, such as route delay and packet loss ratio.

A Greedy-based Service Orientation Algorithm (GSOA) was introduced by [68] to solve the server overload by selecting the closest and compliant Service Functions (SF). The proposed GSOA shows promising results in reducing data transmission time and balancing the SFs load. However, this approach cannot provide efficient load balance due to the proportion of fixed-packets or non-fixed-packets variation. Also, the researchers did not consider QoS realistic factors in the evaluation process. While in [69] the Karush□Kuhn□Tucker conditions were employed to pick the OpenFlow□enabled devices that generate the most load on the controller and pass the fewest users through it. The method improved the end users' QoS metrics by reducing jitter, delay, throughput, and packet loss. However, using KKT alone to minimize the response time could cause irregular load distribution as it depends on the arithmetic configurations of the controllers.

Mean-field game theory and Dijkstra's method were implemented by [70], [71], [72] to optimize resource and bandwidth usage by determining the shortest path to the destination and improving QoS performance. Also, these methods could efficiently mitigate the bottleneck, packet loss ratio, overhead and delay. But it needs more accurate traffic prediction in the case of IoT and satellite contexts.

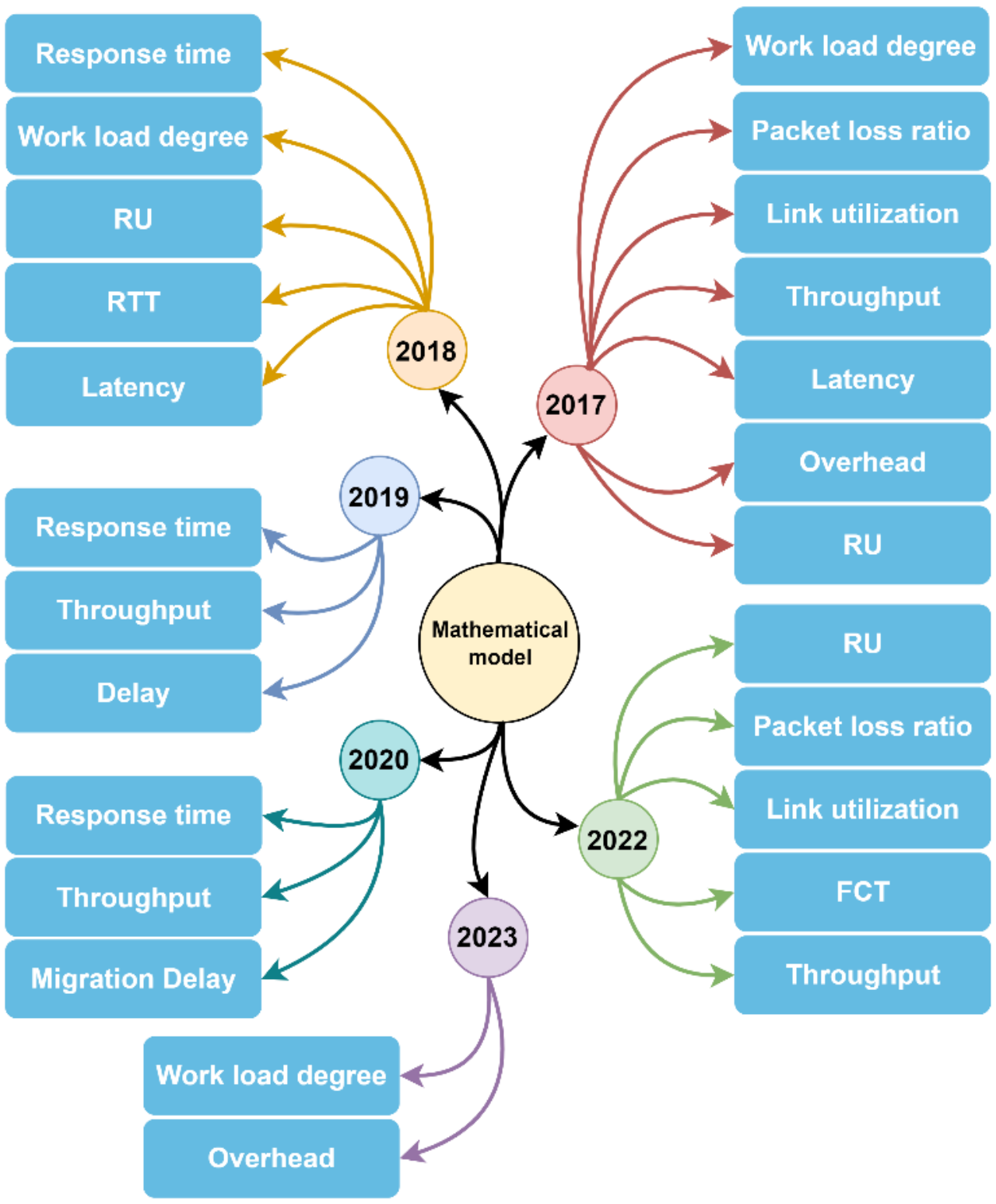

*Figure 7: The metrics trends used by mathematical models-based LB*

Figure 7 highlights the metrics trends that are significantly used from 2017 to 2023 by the LB methods, which adopted mathematical models. Most of the papers used the following parameters: response time, throughput, work load degree, and latency. The majority of the papers used from two to four metrics in the evaluation process. However, parameters such as RU, FCT, migration delay, and over load have been used in a few works, also other QoS parameters are not considered in the surveyed works.

Other LB techniques proposed by [73] considered the response time a critical factor in the path selection process. In this method, the controller collects the response times of each server and chooses the one with the shortest or most consistent response time. This strategy is considered more cost-effective than conventional alternatives due to the decreased hardware needs and software customization. Although it solves many LB issues, this strategy needs to consider ways to reduce energy consumption in server LB.

Algorithms based on statistical flow introduced by [74], [75], [76], [77] applied dynamic LB to several service types in cloud-based SDN. It enhanced the server's CPU efficiency, memory usage, and response time. Despite its advantages over other approaches, this technique has a significant flaw: it relies only on reactive flow entry and does not undertake real-time load monitoring. Services Orchestration and Data Aggregation method (SODA) applied by [78] to overcome the problem of data redundancy and slow service response through cloud-based SDN. However, other QoS indicators are not considered.

In, [79], [80], [81] a dynamic LB method is proposed Based on current computational features and network resource information. This approach depends on different characteristics to select the best path dynamically. The technique enhances the QoS parameters, making the network more stable and effective. The lack of QoS factors, such as cost optimization and network latency, is a significant flaw in this approach.

Switch migration mechanisms proposed by [82], [83], [84], [85], [86] to solve the LB issues through selecting the migrated switch, target controller and migration of switch. Migration techniques used by many researchers as an LB method in SDN, it belongs to the deterministic category. These approaches can accomplish the optimal solution, but the runtime and latency in the large-scale network will increase. Moreover, this technique's primary issue is loading fluctuation, which could occur if the target controller and migration switch are incorrectly identified. Heterogeneous networks have been used by [87]; the method improves the load distribution and reduces the response time. However, the minimum bandwidth for the data transport channel is not guaranteed. The article by [88] suggested a technique for wireless sensor networks termed perceptually important points-based data aggregation (PIP-DA). This strategy aims to reduce the quantity of data readings transmitted, resulting in less energy consumption and a longer lifespan for the network. However, an additional dynamic segmentation algorithm can be utilized at both the sensor node and gateway levels to predict the missing data at the CH and improve proposed method. Also, the authors of [89] introduced a method called "DAEP" for conserving energy in WSNs by aggregating data performed at the individual sensor node level based on extracting extrema points. The proposed technique aims to reduce energy consumption and prolong the lifespan of the WSN. Unfortunately, AI, ML, and statistical techniques were not utilized, which could extend the WSN lifespan. Finally, a centralized network clustering strategy used by [90] in which the base station (BS) splits the network into clusters and identifies the node with the most significant energy level as the cluster head (CH) for each cluster at the start of the protocol.

Figure 8 graphically presents the metrics used by the fourth category to check the effectiveness of the proposed techniques. The majority of articles published between 2017 and 2022 employed response time, throughput, workload degree, and packet loss ratio to demonstrate efficiency. Although many different indicators of quality of service (QoS) should be considered, the articles included in this review only employed one to four QoS metrics to demonstrate the effectiveness of the suggested algorithms.

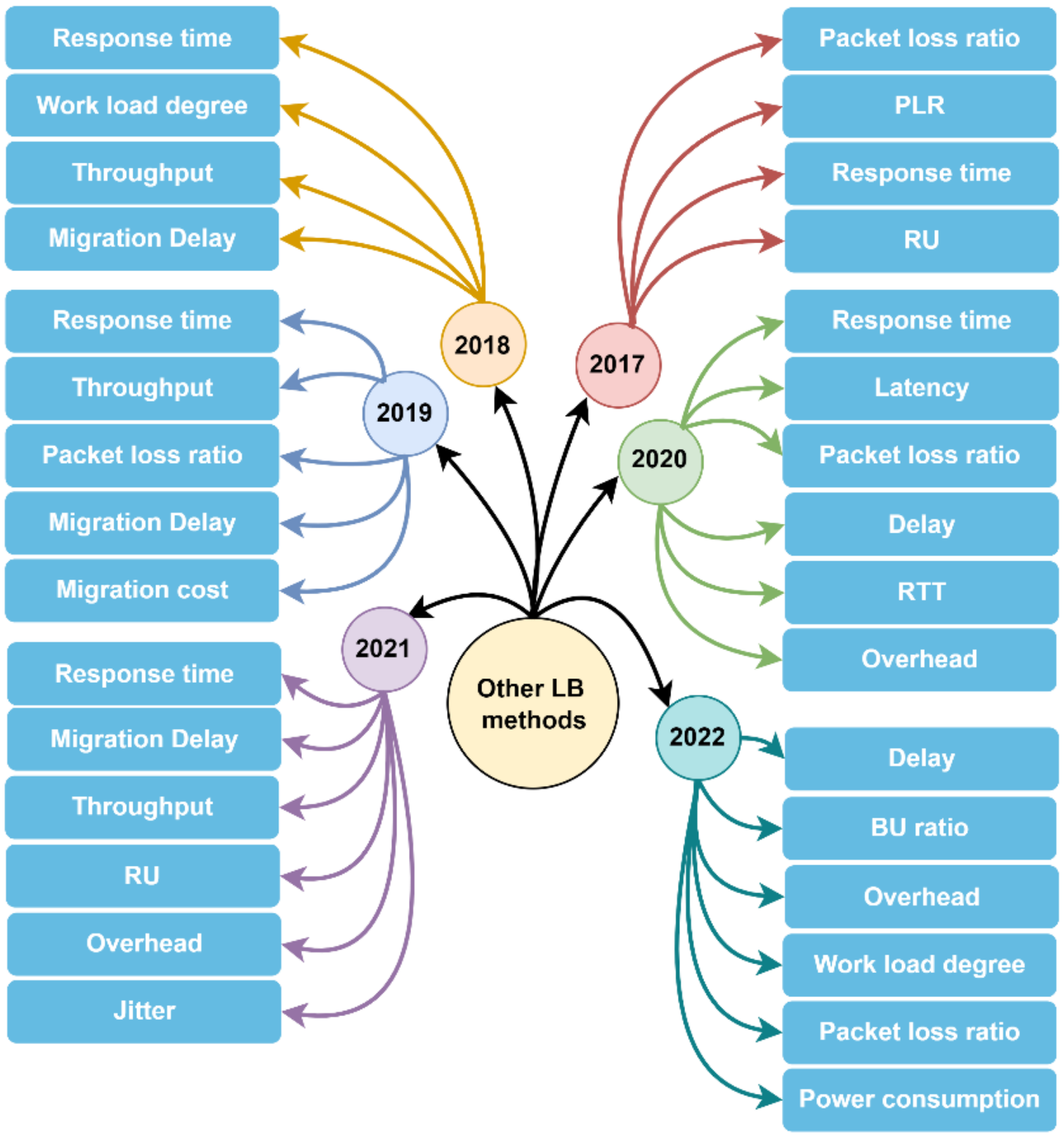

*Figure 8: The metrics trends used by other LB methods*

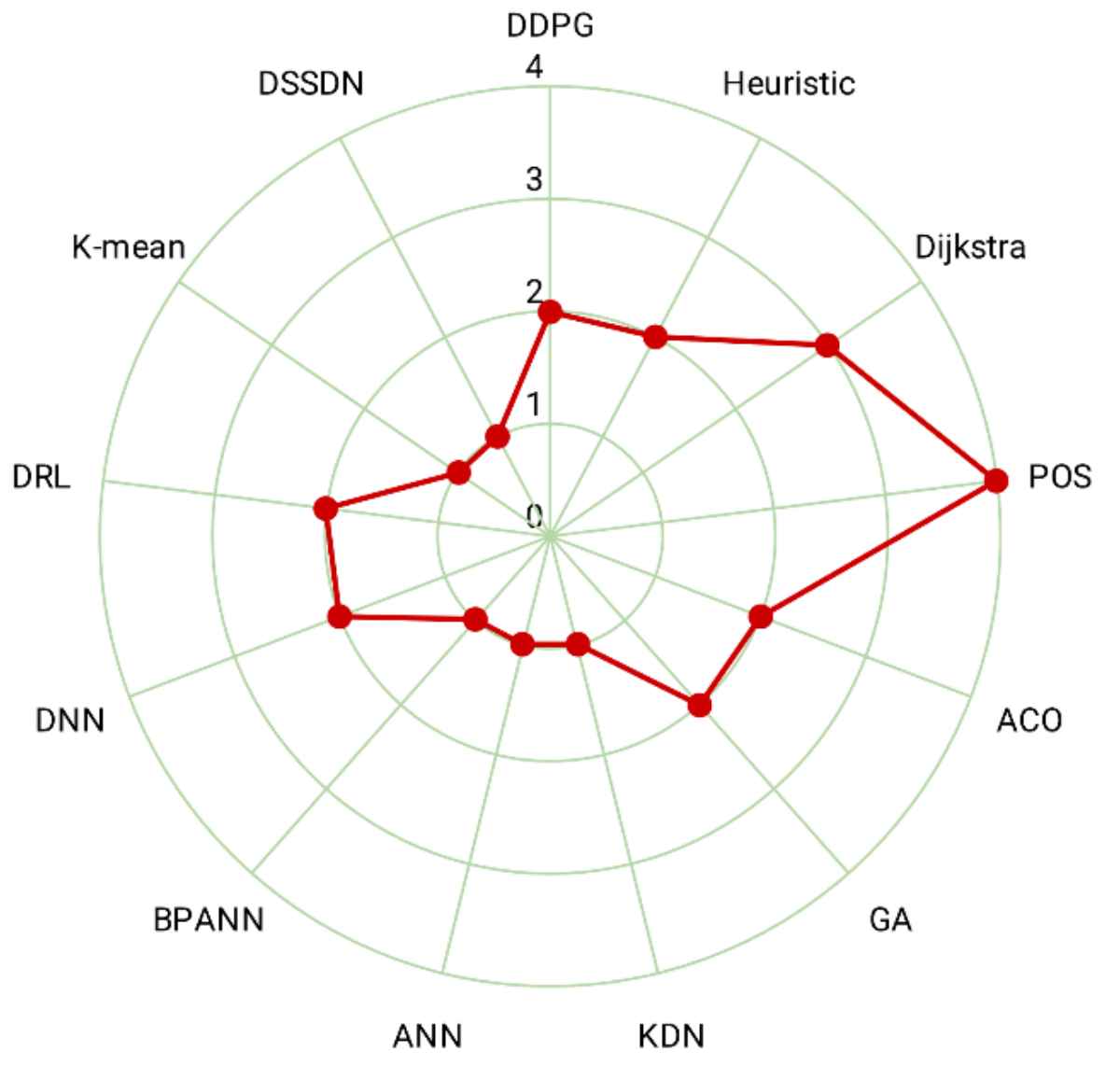

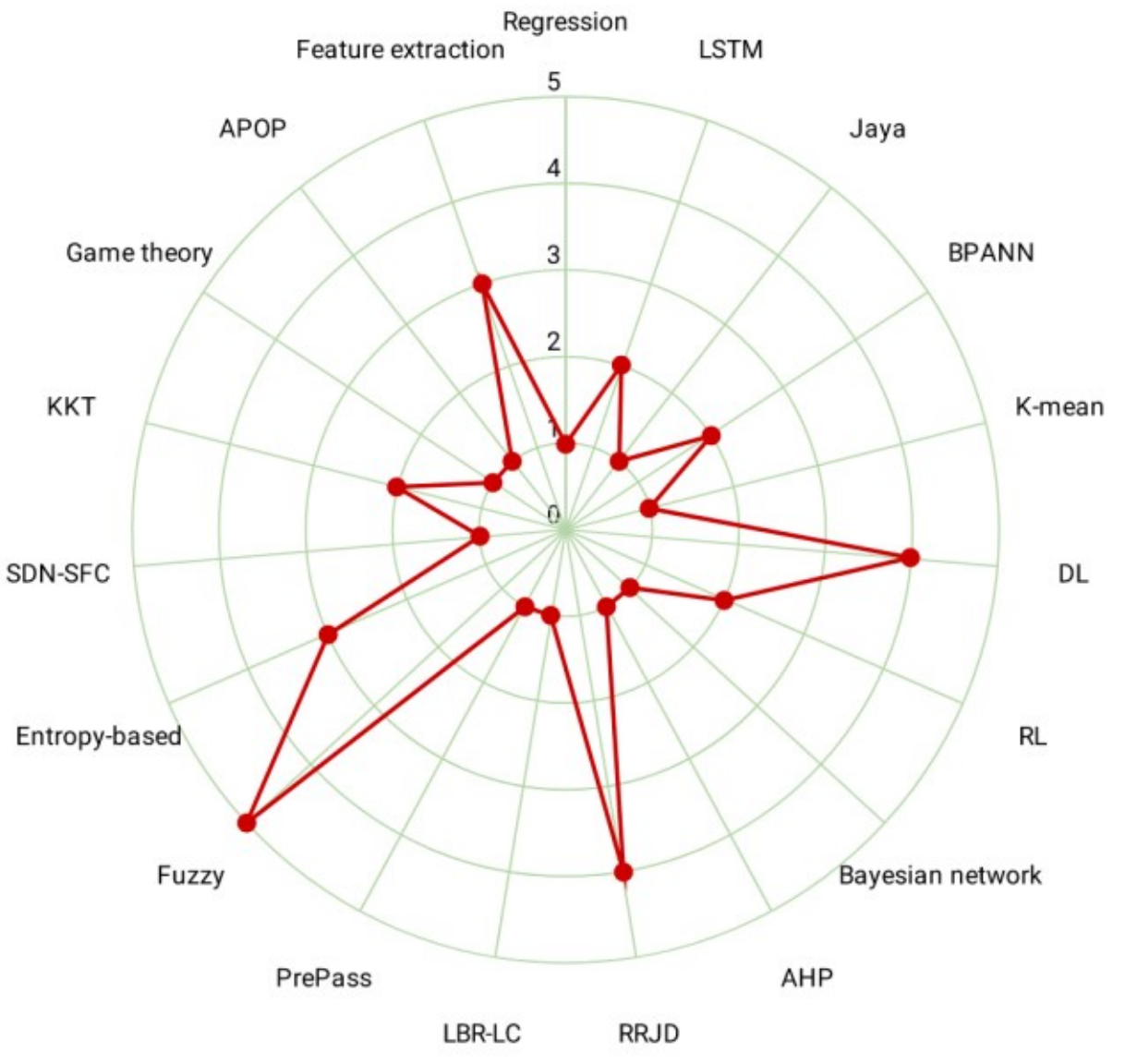

*(a) Optimal path selection methods* *(b) Overload detection algorithms/techniques*

*Figure 9: The algorithms/techniques used in the reviewed papers*

In the proposed LB solutions, different SDN architectures have been suggested, such as centralized controller, distributed but logically centralized controller, and distributed controller. There are advantages and disadvantages associated with each architecture. Scalability issues emerged, for instance, with a centralized single controller. The multiple controllers-based architectures improved the control plane scalability; however, it suffers from uneven load distribution; while some controllers are overloaded, others are unused. In order to devise a successful load management strategy, it is essential to employ the correct load-balancing algorithm for each design. Detailed analysis of the methods used in the surveyed papers to choose the optimal path and the algorithms/techniques applied to detect the network overload are given in figures 9 (a) and (b).

## 5. TRENDS AND CHALLENGES

Critical trends and challenges regarding SDN load balancing have yet to be examined exhaustively and thoroughly. This section lists and discusses some of these trends and challenges obtained from reviewed papers.

1. A single controller failure can be overcome by migration to another controller. However, there is still an overhead associated with moving a load of a failing controller to an operating one. Distributing the controllers into several domains could be a good solution, but the controller deployment cost will increase, and the controller efficiency to handle network change will decrease. In the currently proposed methods, the reduction of load migration overhead and fee has yet to be considered. This issue can be explored in the future, as the failure of a centralized controller could lead to the collapse of the entire network.
2. Initially, SDN was implemented on small enterprise networks. Recently, researchers have proposed several load-balancing methods for medium and small-sized networks. However, applying load balancing for dynamic traffic load failure scenarios in large-scale networks is still a research topic. Therefore, researchers must investigate different load-balancing techniques to build a reliable and effective network.
3. Researchers have relatively limited studies in the field of load balancing utilizing methods informed by artificial intelligence. Instead, they could efficiently employ a hybrid strategy, which combines two or more approaches, as a future path for balancing the load.

4. Most of the surveyed research has yet to show the effect of the used LB mechanism on all QoS parameters. For instance, many methods prioritize factors like throughput, scalability, and response time while overlooking latency, packet loss, and stability parameters. Therefore, load-balancing decision-making must incorporate additional QoS criteria. Furthermore, further research may find it interesting to study global QoS compliance.
5. Some of the chosen papers did not consider factors such as traffic patterns and packet priority. Therefore, using these considerations in load-balancing decisions might be a potential avenue for future study.
6. The energy saving and carbon emission challenge do not consider in the most studied techniques. These factors could increase the popularity and efficacy of existing load-balancing mechanisms. Consequently, load-balancing methods that consider carbon emissions and energy usage are promising research direction.
7. Furthermore, some of the reviewed methods do not incorporate the algorithm used to accomplish load detection. Hence, introducing a novel approach for load detection is another route for future work.
8. Security challenges in SDN are a significant issue that all researchers should considered. The majority of SDN security risks target the availability of the control plane. However, using distributed controllers requires more cost, and it can cause controller cascade failures. Therefore, SDN security must be associated with implementing a secure design that assures control plane high availability.

## CONCLUSION

Using multiple load-balancing techniques in SDN networks can boost network performance since the SDN controller has the capability to provide a comprehensive overview of the available resources. This article provides a detailed and comprehensive survey of different load balancing techniques that adopt artificial intelligence to improve the load distribution in Software Defined Networks. We discuss the SDN architecture, its advantages, and its LB mechanisms. Followed by revising the artificial intelligence-based load-balancing methods researchers proposed regarding implementation and evaluation metrics. Additionally, the paper categorized the current load-balancing solutions for SDN into four primary classifications and explained their applications, each with sub-classifications based on the utilized technology. These classifications include; Nature-inspired LB methods that resemble or are inspired by natural events; Machine learning methods that conjunction with the SDN architecture to achieve enhanced routing performance; Mathematical model-based techniques that use to perform load balancing in software-defined and other methods that apply different ways to predict the network overload and optimal path. We provided detailed information about various metrics associated with the performance evaluation of load balancing in SDN. These metrics include response time, throughput, resource utilization, latency, workload degree, deployment cost, jitter, packet loss ratio, delay, round trip time, bandwidth utilization ratio, migration delay, link utilization, flow completion time, migration cost, overhead, packet load ratio, power consumption, consumer Satisfaction, and cumulative distribution function. Furthermore, we summarized and compared the techniques applied by the surveyed articles and analyzed the limitation of each one. In conclusion, we list and discuss some trends and challenges in potential areas for future investigation that can enhance the widespread adoption of load balancing in SDN. In our future work, we will consider more databases, journals, and conferences. We will also employ more keywords and search strings to search the literature. This review did not incorporate articles published prior to 2017. Additionally, we will cover other related issues, as this work focuses only on using AI in SDN load balancing problems.